\documentclass[ reprint, prb,superscriptaddress,tightenlines,amsmath,amssymb, aps]{revtex4-1}
\usepackage{graphicx}% Include figure files
\usepackage{dcolumn}% Align table columns on decimal point
\usepackage{amsthm}
\usepackage{bm}% bold math

\pdfoutput=1

\usepackage{graphicx,graphics,epsfig,subfigure,times,bm,bbm,amssymb,amsmath,amsthm,mathrsfs,MnSymbol}
\usepackage{gensymb}
\usepackage{amsfonts}
\usepackage[matrix,frame,arrow]{xypic}
\usepackage[pdfstartview=FitH,colorlinks=true,linkcolor=red,citecolor=magenta,filecolor=magenta,urlcolor=cyan,runcolor=cyan]{hyperref}
\usepackage{epstopdf}
\usepackage[pdftex]{color}
\usepackage{braket}%Dirac Notation in QM
\usepackage{enumerate}
\usepackage[normalem]{ulem}
\usepackage[usenames,dvipsnames]{xcolor}
\usepackage{multirow}
\usepackage{mathtools}

\definecolor{orange}{rgb}{1,0.5,0}

\newcommand{\ignore}[1]{}
\usepackage{geometry}

\begin{document}

\title{Probing the dynamical phase transition with a %programmable
superconducting quantum simulator}

\author{Kai Xu}
\thanks{Those authors contributed equally to this work.}
\affiliation{Institute of Physics, Chinese Academy of Sciences, Beijing 100190, China}

\author{Zheng-Hang Sun}
\thanks{Those authors contributed equally to this work.}
\affiliation{Institute of Physics, Chinese Academy of Sciences, Beijing 100190, China}

\author{Wuxin Liu}
\thanks{Those authors contributed equally to this work.}
\affiliation{Interdisciplinary Centre for Quantum Information and Zhejiang Province Key
Laboratory of Quantum Technology and Device, Department of Physics, Zhejiang University,
Hangzhou 310027, China}

\author{Yu-Ran Zhang}
\affiliation{Theoretical Quantum Physics Laboratory, RIKEN Cluster for Pioneering Research, Wako-shi, Saitama 351-0198, Japan}
%\affiliation{Beijing Computational Science Research Centre, Beijing 100094, China}

\author{Hekang Li}
\affiliation{Institute of Physics, Chinese Academy of Sciences, Beijing 100190, China}
\affiliation{Interdisciplinary Centre for Quantum Information and Zhejiang Province Key
Laboratory of Quantum Technology and Device, Department of Physics, Zhejiang University,
Hangzhou 310027, China}

\author{Hang Dong}
\affiliation{Interdisciplinary Centre for Quantum Information and Zhejiang Province Key
Laboratory of Quantum Technology and Device, Department of Physics, Zhejiang University,
Hangzhou 310027, China}

\author{Wenhui Ren}
\affiliation{Interdisciplinary Centre for Quantum Information and Zhejiang Province Key
Laboratory of Quantum Technology and Device, Department of Physics, Zhejiang University,
Hangzhou 310027, China}

\author{Pengfei Zhang}
\affiliation{Interdisciplinary Centre for Quantum Information and Zhejiang Province Key
Laboratory of Quantum Technology and Device, Department of Physics, Zhejiang University,
Hangzhou 310027, China}

\author{Franco Nori}
\affiliation{Theoretical Quantum Physics Laboratory, RIKEN Cluster for Pioneering Research, Wako-shi, Saitama 351-0198, Japan}
\affiliation{Physics Department, University of Michigan, Ann Arbor, MI 48109-1040, USA}

\author{Dongning Zheng}
\email{dzheng@iphy.ac.cn}
\affiliation{Institute of Physics, Chinese Academy of Sciences, Beijing 100190, China}
\affiliation{CAS Centre for Excellence in Topological Quantum Computation, School of Physical Sciences, UCAS, Beijing 100190, China}

\author{Heng Fan}
\email{hfan@iphy.ac.cn}
\affiliation{Institute of Physics, Chinese Academy of Sciences, Beijing 100190, China}
\affiliation{CAS Centre for Excellence in Topological Quantum Computation, School of Physical Sciences, UCAS, Beijing 100190, China}

\author{H. Wang}
\email{hhwang@zju.edu.cn}
\affiliation{Interdisciplinary Centre for Quantum Information and Zhejiang Province Key
Laboratory of Quantum Technology and Device, Department of Physics, Zhejiang University,
Hangzhou 310027, China}

\begin{abstract}
Non-equilibrium  quantum many-body systems, which are difficult to study via classical computation, have attracted wide interest. Quantum simulation can provide insights into these problems. Here, using a programmable quantum simulator with 16 all-to-all connected superconducting qubits, we investigate the dynamical phase transition in the Lipkin-Meshkov-Glick model
with a quenched transverse field. Clear signatures of the dynamical phase transition, merging different concepts of dynamical criticality, are observed by measuring the non-equilibrium order parameter, nonlocal correlations, and the Loschmidt echo. Moreover, near the dynamical critical point, we  obtain the optimal spin squeezing of $-7.0\pm 0.8$ decibels, showing  multipartite entanglement useful for   measurements with precision five-fold beyond the standard quantum limit. Based on the capability of entangling qubits simultaneously and the accurate single-shot readout of multi-qubit states, this superconducting quantum simulator can be used to study other problems in non-equilibrium quantum many-body systems.
\end{abstract}
\maketitle

\section{Introduction}
Quantum simulation uses a controllable quantum system to mimic complex systems or solve intractable problems~\cite{ref1,ref2}. Emergent phenomena in out-of-equilibrium quantum many-body systems~\cite{ref3}, e.g., thermalization~\cite{ref4} versus localization~\cite{ref5}, and time crystals~\cite{ref6}, have attracted considerable attentions using quantum simulation.
Recently, the dynamical phase transition (DPT) and the non-equilibrium phase transition in transient time scales have been theoretically studied in the transverse-field Ising model with all-to-all interactions~\cite{ref7,ref8,ref9}. These two transitions can be characterized by a non-equilibrium order parameter~\cite{ref7,ref8,ref9,ref10}, and the Loschmidt echo associated with the Lee-Yang-Fisher zeros in statistical mechanics~\cite{ref11}, respectively. Moreover, recent experimental progress has allowed for the controllable simulation of these exotic phenomena with cold atoms~\cite{ref12,ref13} and trapped ions~\cite{ref14,ref15}. Yet, experimental explorations for the dynamics of entanglement, as a valuable resource in quantum information processing, remain limited in the presence of a DPT.

In our experiments, applying a sudden change of the transverse field with a controllable strength, we drive the system, initially in its ground state, out of equilibrium. Accurate single-shot readout techniques enable us to synchronously record the dynamics of all qubits and to observe
essential signatures of DPTs and spin squeezing from the dynamical criticality in the Lipkin-Meshkov-Glick (LMG) model.

This work presents a systematic quantum simulation of DPTs with two different concepts, providing an evidence of the relation between the non-equilibrium order parameter and the Loschmidt echo. More importantly, we verify entanglement in spin-squeezed states generated from the dynamical criticality, directly observing squeezing of $-7.0\pm 0.8$ decibels for 16 qubits. 

\section{Results}
%%%%%%%%%%%%%%%%%%%%%%%%%%%%%%%%%%%%%%%%%%%%%%%%%%%%%%%%%%%%%%%%%%%%%%%
\begin{figure*}\centering
	\includegraphics[width=0.7\linewidth]{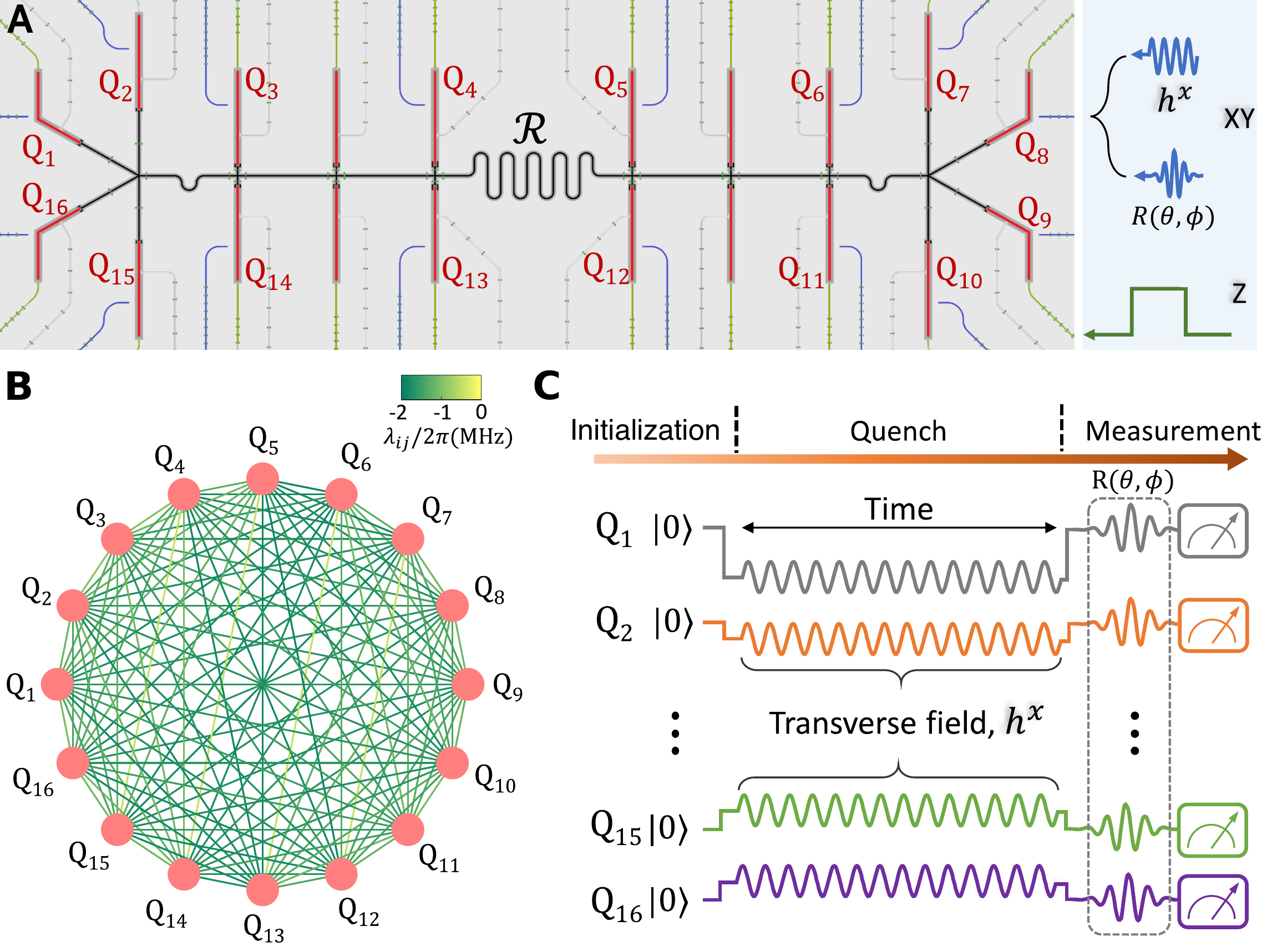}
	\caption{\textbf{Quantum simulator and experimental pulse sequences.} \textbf{A}, Optical micrograph of the device with colours added in different regions to distinguish qubits (red), the resonator bus (black), qubit XY-control lines (blue), and Z-control lines (green), respectively. All qubits are capacitively coupled to the resonator bus ($\mathcal{R}$). Each qubit $Q_j$ can be individually controlled by its own Z-control line for the frequency tuning. \textbf{B}, Connectivity graph of the 16-qubit system when all qubits are equally detuned from the resonator bus by $\Delta/2\pi\simeq-450$~MHz, with the coloured straight lines representing the magnitude of the qubit-qubit couplings. \textbf{C}, The experimental pulse sequences for simulating the DPT. The sequences are divided into three steps. First, the qubits are initialised at the $\vert 00...0\rangle$ state at their corresponding idle frequencies.
Then, rectangular pulses are applied to quickly bias each qubit to the same detune from the common resonator bus, to turn on the qubit-qubit interactions. Almost simultaneously, the system is quenched by driving the qubits with resonant microwave pulses for a time $t$.
Finally, the qubits are biased back to their idle frequencies before the 16-qubit joint readout is executed, yielding the probabilities \{$P_{00...0}, P_{00...1}, ..., P_{11...1}$\}, from which $\langle\sigma^z_j\rangle$ can be calculated. When necessary, single-qubit rotation pulses $R_j(\theta_j,\phi_j)=\exp[-i\hbar\theta_j(\cos\phi_j\sigma^x_j+\sin\phi_j\sigma^y_j)/2]$ (in black dotted box) are applied in advance to bring the axis defined by $(\theta_j,\phi_j+\pi/2)$ in the Bloch sphere of $Q_j$ to the $\sigma^z$ direction before the readout.}\label{fig1}
\end{figure*}

Our quantum simulator is a superconducting circuit with 20 fully-controllable transmon qubits capacitively coupled to a resonator bus $\mathcal{R}$ (Fig.~1A). Sixteen qubits ($Q_1$--$Q_{16}$), %having  their
with XY-control lines,
are selected to perform  experiments (see Materials and Methods). The resonant frequency of $\mathcal{R}$ is fixed at about 5.51~GHz, while the qubit frequencies are individually tuneable via their Z-control lines, enabling us to engineer the qubit-qubit interactions induced by $\mathcal{R}$. We detune all 16 qubits %($Q_1$-$Q_{16}$)
from $\mathcal{R}$ by, %$\Delta$,
e.g., $\Delta/2\pi\simeq -450$~MHz, to switch on the resonator-mediated interactions between two arbitrary  qubits~\cite{ref16}. Simultaneously, identical resonant microwave drives, with a magnitude of $h^{x}$, are imposed on all qubits   to generate the local transverse fields for the control of a DPT (Fig.~1C). The crosstalk effects of microwave pulses have been precisely corrected to ensure the uniformity of the local fields (see Supplementary Materials). The effective Hamiltonian of the quenched system is
\begin{equation}
H_{1}/\hbar = \sum_{i\neq j}^{N}\lambda_{ij}(\sigma_{i}^{+}\sigma_{j}^{-}+\sigma_{i}^{-}\sigma_{j}^{+}) + h^{x} \sum_{j=1}^{N}\sigma_{j}^{x},
\label{H1}
\end{equation}
where $N=16$, $\lambda_{ij}\equiv g_{i}g_{j}/\Delta+\lambda_{ij}^{c}$ is the qubit-qubit coupling
strength, $g_j$ represents the coupling strength between $\mathcal{R}$ and $Q_j$,
 $g_{i}g_{j}/\Delta$ is the resonator-induced virtual coupling strength between $Q_i$ and $Q_j$, which
 is much larger than the crosstalk coupling $\lambda_{ij}^c$ (Parameters are
 shown in Supplementary Materials).  %The parameters of the quantum simulator are shown in Extended Data Table~1.
Since the values of $\lambda_{ij}$ are nearly the same for most pairs of qubits and
do not decay over a distance $\vert i-j\vert$ %, i.e., $\lambda_{ij}$ is set to be uniform
(Fig.~1B), the quenched system can be reasonably approximated by the LMG model, whose Hamiltonian
is $H_{\text{LMG}} = (J/N) (\mathcal{S}^{z})^{2} + \mu \mathcal{S}^{x}$, with $\mathcal{S}^{x,z}\equiv \sum_j\sigma_j^{x,z}/2$ %as derived from Eq.~(1)
(see Materials and Methods). Recent studies~\cite{ref7,ref8,ref9,ref10,ref14} have shown that $H_{\text{LMG}}$ has a
dynamical critical point $\mu/J=1/2$, separating the  dynamical paramagnetic phase (DPP) and
the dynamical ferromagnetic
phase (DFP) with and without a global $\mathbb{Z}_{2}$ symmetry, respectively.
%The dynamical ferromagnetic phase
%has a nonzero non-equilibrium order parameter without a global $\mathbb{Z}_{2}$
%symmetry, while the dynamical ferromagnetic phase has a zero non-equilibrium
%order parameter %with a zero value
%and restores the symmetry.

%%In addition, the DPT can also be characterised by the local minimum of two-spin correlations \cite{ref16}.

%By using our programmable quantum simulator,
%we are able to simulate and verify those phenomena. The
%measured results match considerably well with theoretical
%expectations.

\begin{figure}\centering
	\includegraphics[width=0.99\linewidth]{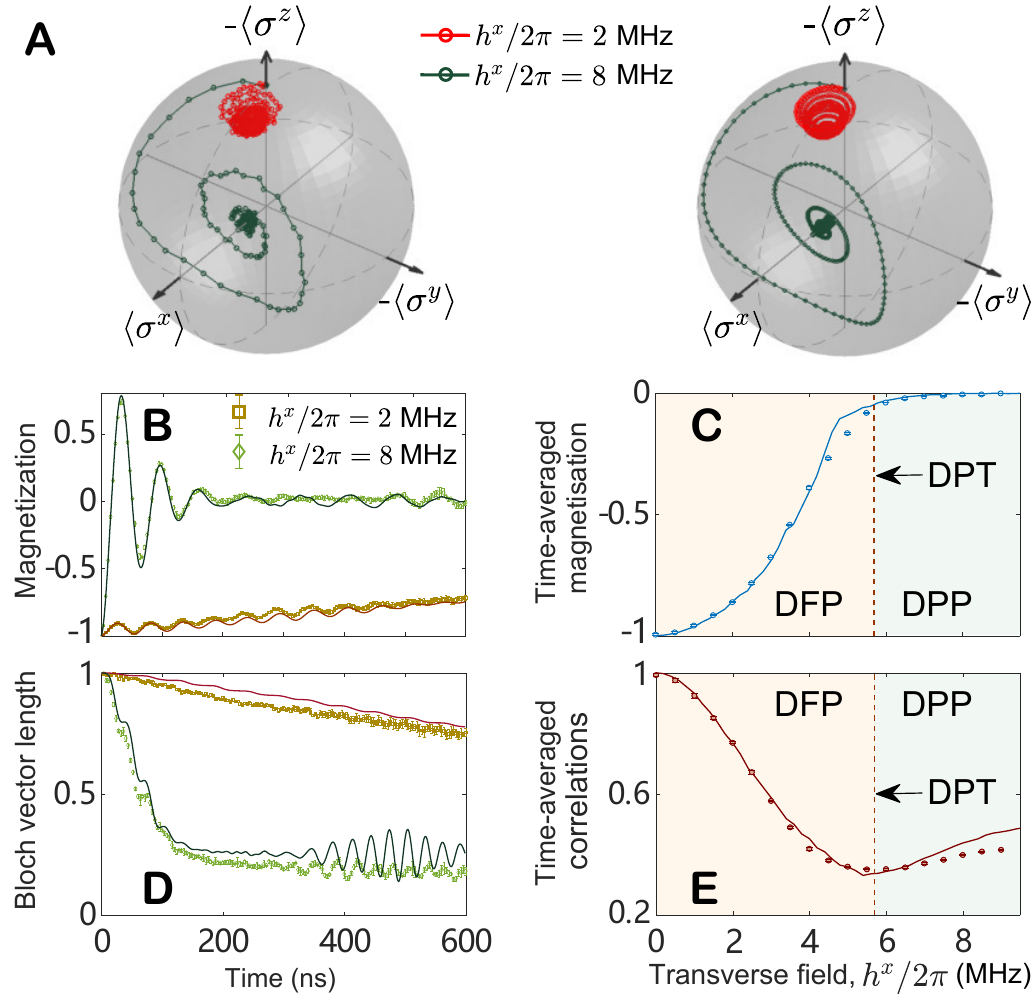}
	\caption{\textbf{Magnetisation and spin correlation.} \textbf{A,} Experimental and numerical data of the time evolution of the average spin magnetisation shown in the Bloch sphere for different strengths of the transverse fields. \textbf{B,} The time evolution of the magnetisation $\langle\sigma^{z}\rangle(t)$. \textbf{C,} The nonequilibrium order parameter, $\overline{\langle\sigma^{z}\rangle}$, as a function of $h^{x}/2\pi$. \textbf{D,} The dynamics of the Bloch vector length $|\langle \vec{\sigma} \rangle|$. \textbf{E,} The averaged spin correlation $\overline{C_{zz}}$ versus $h^{x}/2\pi$. The regions with light red and light blue in \textbf{C} and \textbf{E} show the dynamical ferromagnetic phase and dynamical paramagnetic phase, respectively, separated by a theoretically predicted critical point $h_{c}^{x}/2\pi\simeq 5.7$~MHz. The solid curves in \textbf{B--E} are the numerical results using Eq.~(1) without considering decoherence.}\label{fig2}
\end{figure}

\begin{figure}\centering
	\includegraphics[width=0.99\linewidth]{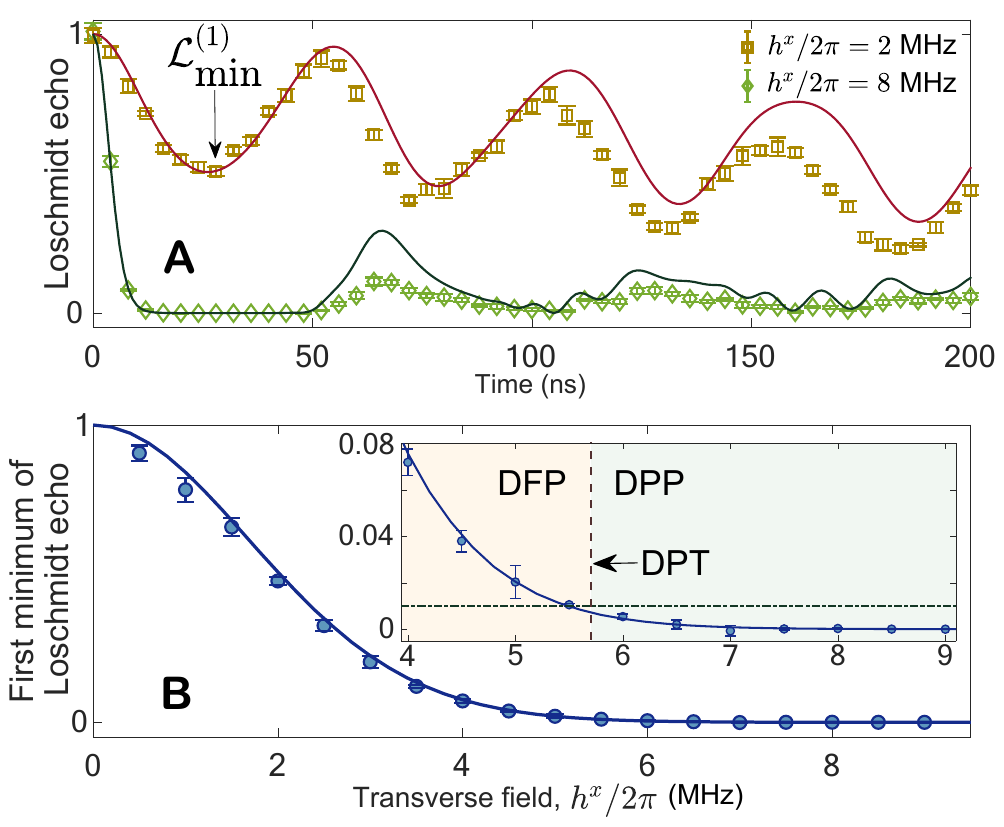}
	\caption{\textbf{Loschmidt echo.} \textbf{A,} Time evolution of the Loschmidt echo $\mathcal{L}(t)$ for different transverse field strengths. \textbf{B,} The earliest minimum point of $\mathcal{L}(t)$ during its dynamics, $\mathcal{L}_{\text{min}}^{(1)}$, as a function of $h^{x}$. The behaviour of $\mathcal{L}(t)$ for longer time indicates the existence of anomalous dynamical phases \cite{ref12} (see Methods). Additionally, we note that the Loschmidt echo cannot be strictly equal to 0, because of  finite-size effects. We demonstrate that
$\mathcal{L}_{\text{min}}^{(1)}$ reaches $\sim10^{-2}$ at the transition point and becomes smaller in the paramagnetic phase for the LMG model with $N=16$ (see Methods).
%The regions with light yellow and light green in \textbf{b} show the dynamical ferromagnetic and paramagnetic phases, respectively.
The solid curves in \textbf{A} and \textbf{B} are the numerical results using  Eq.~(1) without considering decoherence.}\label{fig3}
\end{figure}

First, we show that our programmable superconducting qubits can simulate and verify the
DPT
%, matching the theoretical expectation,
by measuring the magnetisation and the spin correlation.
%, which match
%well with theoretical expectations.
The system is initialised at the eigenstate $|00\ldots 0\rangle$ of $H_1$ with $h^x=0$, where $|0\rangle$ denotes the ground state of a qubit. Then, we quench the system by suddenly adding a transverse field and monitor its dynamics from the %at different evolution
time $t=0$  to 600~ns. %(0-600~ns).
With the precise full control and the high-fidelity single-shot readout of each qubit, we are able to omni-directionally track the evolutions of the average magnetisation
\begin{equation}
\langle\sigma^{\alpha}(t)\rangle\equiv\frac{1}{N}\sum_{j=1}^{N} \langle\sigma_{j}^{\alpha}(t)\rangle, \nonumber
\label{op}
\end{equation}
along the $x$,$y$,$z$-axes for different strengths of the quenched transverse fields, with $\alpha \in \{x,y,z\}$. %\langle\psi(t)|\psi(t)\rangle$ and $|\psi(t)\rangle=e^{-iH_{1}t/\hbar}|00\ldots 0\rangle$.
By depicting the trajectory of the Bloch vector
$\langle \vec{\sigma} \rangle = [\langle\sigma^{x}\rangle,~\langle\sigma^{y}\rangle,~\langle\sigma^{z}\rangle$],
the dynamics of our quantum simulator with two distinct transverse fields
is visualised in Fig.~2A.
%on the Bloch sphere.
For a small transverse field, e.g.,
$h^{x}/2\pi\simeq 2$~MHz,
%the spin alignment preserves the symmetry-broken ferromagnetic  state
%during the time evolution, i.e.,  %.
% when applying small transverse fields ($h^{x}/2\pi\simeq 2$~MHz).
%In this case,
  $\langle\sigma^{z}(t)\rangle$ exhibits a slow relaxation (Fig.~2B). However, given
  a strong transverse field, e.g., $h^{x}/2\pi\simeq 8$~MHz, %the spin alignment changes to a paramagnetic state, i.e.,
  $\langle\sigma^{z}(t)\rangle$ exhibits a large oscillation at an early time and approaches  zero
  in the long-time limit
(Fig.~2B). In Fig.~2C, we show the behaviour of the time-averaged magnetisation,
$\overline{\langle\sigma^{z}\rangle}\equiv({1}/{t_{f}})\int_{0}^{t_{f}}\!\!dt\langle\sigma^{z}(t)\rangle$,
that is defined as the non-equilibrium order parameter. It demonstrates that $\overline{\langle\sigma^{z}\rangle}\neq 0$ and $\overline{\langle\sigma^{z}\rangle}= 0$ in the DFP and the DPP, respectively. The experimental data of $\overline{\langle\sigma^{z}\rangle}$ for qubits with different detunings $\Delta$ are presented in Supplementary Materials. In addition, the Bloch vector length $|\langle \vec{\sigma} \rangle|$ also depends on the strength of the transverse field $h^{x}$.
For large $h^{x}$, $|\langle \vec{\sigma} \rangle|$ decays rapidly to a small value,
indicating strong quantum fluctuations in the DPP~\cite{ref11} (Fig.~2D).
Figure~2E  shows the averaged spin correlation function,
$\overline{C_{zz}}\equiv ({1}/{t_{f}})\int_{0}^{t_{f}}\!\!dt\sum_{ij} \langle\sigma_{i}^{z}(t)\sigma_{j}^{z} (t)\rangle/N^2$,
versus $h^x$ with a final time $t_f=600$~ns, where the DPT is characterised by the local minimum of two-spin correlations \cite{ref16}. %From
%the experimental results,
%$\overline{\sigma^{z}}$ increases monotonously from $-1$ to 0, as $h^x$ increases.
We can observe the critical behaviours of $\overline{\langle\sigma^{z}\rangle}$ and
$\overline{C_{zz}}$ as the signatures of the DPT, when the transverse field strength is set near  %is
%The experimentally-obtained
%critical transverse field strength for this crossover is ,
%in the good agreement with
the theoretical prediction $h^{x}_{c}/2\pi =N |\lambda|/8\pi\simeq5.7$~MHz %($\lambda<0$)
with $\lambda\equiv\overline{\lambda_{ij}}$
%by considering the results of $H_{\text{LMG}}$
(see Materials and Methods). %In addition, the emergence of a dip in
%$\overline{C_{zz}}$ is observed close to the critical point (Fig.~2d), showing a clearer signature of
% DPTs between two dynamical phases~\cite{ref17}.

\begin{figure*}\centering
	\includegraphics[width=0.99\linewidth]{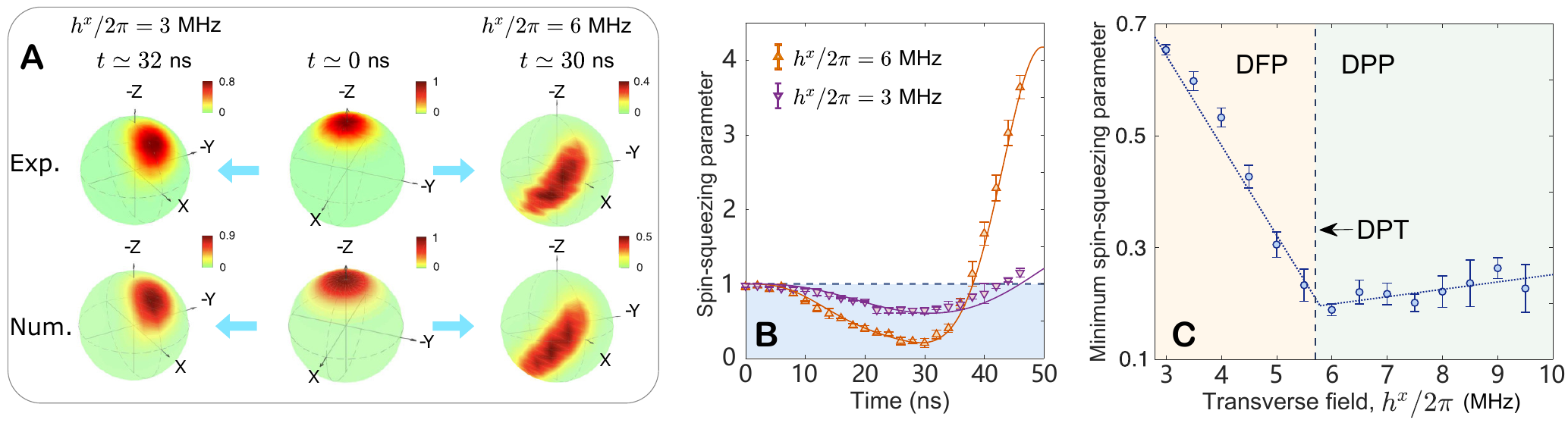}
	\caption{\textbf{Quasidistribution $Q$-function and  spin-squeezing parameter.} \textbf{A,} Experimental and numerical data of $Q(\theta,\phi)$ in the spherical coordinates, when the minimum values of the spin-squeezing parameters are achieved during the time evolutions with the strengths of the transverse fields $h^{x}/2\pi\simeq 3$~MHz and 6~MHz, respectively. \textbf{B,} Time evolution of the spin-squeezing parameters with $h^{x}/2\pi\simeq 3$~MHz and 6~MHz, respectively. \textbf{C,} The minimum spin-squeezing parameter $\xi_{\text{min}}^{2}$ as a function of $h^{x}$. The solid lines in \textbf{B} are the numerical results using Eq.~(1) without considering decoherence. The blue shaded area in \textbf{B} is only accessible for entangled states. The dotted line in \textbf{C} is the piecewise linear fit, whose minimum point is close to the theoretically predicted critical point $h_{c}^{x}/2\pi\simeq 5.7$~MHz (dashed line).}\label{fig4}
\end{figure*}

Another perspective on dynamical criticality is based on the Loschmidt echo, defined as $\mathcal{L}(t)=|\langle 00\ldots 0|e^{-iH_{1}t/\hbar}|00\ldots0\rangle|^{2}$, where %whose zeros, i.e.,
the time $t$, satisfying $\mathcal{L}(t)= 0$, is a Lee-Yang-Fisher zero. The zero will cause the nonanalytical behaviour of the rate function, $r(t)=-N^{-1} \log[\mathcal{L}(t)]$, regarded as  the complex-plane generalisation of the free energy
density~\cite{ref11}. Recent numerical
studies~\cite{ref7,ref9} have revealed that the existence of Lee-Yang-Fisher zeros closely relates to the DPT between the DFP and the DPP in  long-range interacting systems. %Next, we will experimentally demonstrate the relation.
In Fig.~3A, we show distinct behaviours of the Loschmidt echo in different
dynamical phases.
Here, we consider the first minimum  of the Loschmidt echo $\mathcal{L}_{\text{min}}^{(1)}$ (Fig.~3A), which is sufficient
for judging whether the zeros exist or not.
It approaches  zero in the DPP, and remains relatively large in the DFP~\cite{ref9}. In Fig.~3B, we plot
$\mathcal{L}_{\text{min}}^{(1)}$ versus $h^{x}$ to show the relationship between the DPT and the Loschmidt echo. Similar as the numerical results of the LMG model with  finite-size effects (see Methods), we experimentally observe that $\mathcal{L}_{\text{min}}^{(1)}\leq 0.01$ in the DPP, and it becomes
relatively large $\mathcal{L}_{\text{min}}^{(1)}>0.01$ in the DFP. Our work is the first experiment to combine  the
non-equilibrium order parameter and the Loschmidt echo
for verifying a DPT.
%%%%%%%%%%%%%%%%%%%%%%%%%%%%%%%%%%%%%%%%%%%%%%%%%%%%%%%%%%%%%%%%%%%%%%%

%The above results clearly characterise the DPT in the LMG model
%with a transverse field,
In addition to  demonstrating a DPT, the LMG model is also useful for generating the spin-squeezed state with twist-and-turn dynamics~\cite{ref17,ref18}. Near the equilibrium critical point,  spin squeezing can be achieved,  originating from  quantum fluctuations, according to the Heisenberg uncertainty principle~\cite{ref19}. Similarly, we show that  the spin-squeezed state can also be generated from the dynamical criticality. During the dynamics of the quenched Hamiltonian (\ref{H1}), we can visualise the spin-squeezed state by measuring the quasidistribution $Q$-function~\cite{ref20} $Q(\theta,\phi) \propto \langle\theta,\phi|\rho(t)|\theta,\phi\rangle$, where $|\theta,\phi\rangle\equiv\bigotimes_{j=1}^N(\cos\frac{\theta}{2}|0\rangle_j+\sin\frac{\theta}{2} e^{i\phi}|1\rangle_j)$ is the spin coherent state. The measurement is realised by applying a single-qubit rotation to bring the axis defined by ($\theta,\phi$) in the Bloch sphere to the $z$-axis for each qubit before the joint readout. The experimental and numerical data of $Q(\theta,\phi)$ are compared in Fig.~4A, which clearly show  spin squeezing with a large strength of the external field, due to stronger quantum fluctuations in the DPP (see also Fig.~2C).
%
%To quantitatively study spin squeezing,

We also measured the time-evolved spin-squeezing parameter~\cite{ref17} (see Supplementary Materials)
\begin{equation}
\xi^{2}={4{\min}_{\vec{n}_{\perp}}[\text{Var}(\mathcal{S}^{\vec{n}_{\perp}})]}/{N},
\label{sp}
\end{equation}
where $\vec{n}_{\perp}$ denotes an axis perpendicular to the mean spin direction, and $\text{Var}(\mathcal{S}^{\vec{n}_{\perp}})=\langle(\mathcal{S}^{\vec{n}_{\perp}})^2\rangle - \langle\mathcal{S}^{\vec{n}_{\perp}}\rangle^{2}$.
%The detailed measurement protocol for $\xi^{2}$ is presented in Methods.
In Fig.~4B, we show that $\xi^{2}<1$ %~\cite{ref21,ref22,}
 (a sufficient condition for particle entanglement~\cite{ref21,ref22}) occurs in the time interval  $t\lesssim46$~ns when $h^{x}/2\pi\simeq 3$~MHz, and for $t\lesssim38$~ns when $h^{x}/2\pi\simeq 6$~MHz.
%, respectively where $\xi_{S}^{2}<1$ is satisfied showing that the state $|\psi(t)\rangle$ is spin-squeezed~\cite{ref20,ref21,ref22} and particle entangled~\cite{ref26}.
The minimum spin squeezing parameter over time, $\xi_{\text{min}}^{2}$, as a function of $h^{x}$ is shown in Fig.~4C, where the minimum value $\xi_{\text{min}}^{2}\simeq 0.2$ ($-$7.0~dB) is attained very close to the critical point of the DPT.
Compared with the theoretical limit, about $N^{-2/3}$, of the squeezing parameter for an $N$-body one-axis twisting Hamiltonian~\cite{ref21}, our 16-qubit system %($N=16$)%
achieves a spin squeezing parameter satisfying $\xi_{\text{min}}^{2}\simeq N^{-\alpha}$, with $\alpha\simeq 0.58$. It indicates the high-efficiency generation of the spin-squeezed state from  dynamical criticality, and reveals the potential application of the DPT in quantum metrology.

\section{Discussion}
We have presented clear signatures and entanglement behaviors of the DPT in the LMG model with a superconducting quantum simulator featuring all-to-all connectivity, including the non-equilibrium order parameter, the Loschmidt echo, and spin squeezing. Based on its high degree of controllability, precise measurement, and long decoherence time, our platform with all-to-all connectivity is powerful for generating multipartite entanglement~\cite{ref20,ref23} and investigating non-trivial properties of out-of-equilibrium quantum many-body systems, such as many-body localization~\cite{ref24,ref25}, quantum chaos in Floquet systems~\cite{ref26}, and quantum annealing~\cite{ref27}. 

\section{Materials and Methods}
\subsection{Device information and system Hamiltonian}
The device used here consists of 20 frequency-tuneable superconducting qubits capacitively coupled to a central resonator bus. It is the same circuit presented in Ref.~\cite{ref20}, where more details about the device, the qubit manipulation, and the readout can be found. In Table~\ref{table1} (see Supplementary Materials), we present the characteristics for the quantum simulator involving 16 out of the 20 qubits, with XY-control lines, which have been relabelled in the experiments.

The unused four qubits in this device, without XY-control lines, are detuned far off resonance from the other 16 qubits to avoid interacting with them during the experiments. Thus, they will not be included in the following descriptions. The system Hamiltonian, without applying external transverse fields, can be written as
\begin{align}
H_{s1} /\hbar = &\omega_{\mathcal{R}} a^{\dagger}a + \sum_{j=1}^{16}{\left[\omega_{j}(t)|1_j\rangle \langle 1_j|
	+ g_{j}(\sigma _{j}^{+}a+\sigma _{j}^{-}a^{\dagger})\right]}\nonumber \\
&+ \sum_{i<j}^{16} \lambda^\textrm{c}_{ij}(\sigma _{i}^{+} \sigma _{j}^{-} +\sigma _{j}^{-}\sigma_{i}^{+}), \nonumber %\label{eq1}
\end{align}
where $\omega_{\mathcal{R}}$ and $\omega_j$ represent the fixed resonant frequency and the tuneable frequency of $Q_j$, respectively, while $g_j$ is the coupling strength between the $Q_j$ and resonator bus. The magnitude of the crosstalk coupling between $Q_i$ and $Q_j$ beyond the resonator-induced virtual coupling is denoted as $\lambda_{ij}^c$. When equally detuning all the 16 qubits from the resonator bus by about $\Delta/2\pi\simeq -450$~MHz, and simultaneously applying resonant microwaves to each qubit, the system Hamiltonian can be transformed to
\begin{align}
H_{s2} /\hbar = &\sum_{i<j}^{16} (\lambda^\textrm{c}_{ij}+{g_ig_j}/{\Delta})(\sigma _{i}^{+} \sigma _{j}^{-} +\sigma _{j}^{-}\sigma_{i}^{+}) \nonumber\\
&+\sum_{j=1}^{16} h^x_j(\sigma_j^-e^{i\phi_j}+\sigma_j^+e^{-i\phi_j})\nonumber  %\label{eq2}
\end{align}
with $g_ig_j/\Delta$ being the magnitude of the resonator-mediated coupling between $Q_i$ and $Q_j$. It acts as a dominant part of the qubit-qubit interaction terms, because the crosstalk coupling $\lambda^\textrm{c}_{ij}$ is much smaller. In Fig.~1b, we plot the connectivity graph of the total coupling strength $\lambda_{ij}$ for all the combinations of pairs of qubits. The individually-controllable amplitude and the phase of the microwave drive on each $Q_j$  are represented by $h_j^x$ and $\phi_j$, respectively. In our experiments, we set the uniform amplitude and phase for all  qubits, leading to the  Hamiltonian in Eq.~(1) in the main text. To ensure this uniformity, the calibration process for the microwave drives is described in the Supplementary Materials.
\subsection{Relation between the quantum simulator and the LMG model}
The experimental technologies ensure that the device can be described via the Hamiltonian in Eq.~(1) ($H_{1}$) in the main text. With uniform couplings $\lambda\equiv\overline{\lambda_{ij}}$, the first term of Eq.~(1) 
can be written as 
\begin{eqnarray}
\lambda\sum_{i\neq j}^{16}(\sigma_{i}^{+}\sigma_{j}^{-}+\text{H.c.}) = (J/N) [\mathcal{S}^2 - (\mathcal{S}^{z})^2], \nonumber
\label{}
\end{eqnarray}
where $J \equiv N\lambda$. The second term can be directly rewritten as $h^{x}\sum_{i=1}^{16}\sigma_{i}^{x}=
g\mathcal{S}^x$, with $g=2h^x$. According to $[\mathcal{S}^2,\mathcal{S}^\alpha]=0$ $(\alpha\in\{x,y,z\})$, and 
the fact that the initial state $|00\ldots0 \rangle$ is an eigenstate of $\mathcal{S}^{2}$, we have 
\begin{eqnarray}
\exp [-i (H_{1}/\hbar)t]|00\ldots 0 \rangle \propto \exp (-i H_{\text{LMG}}t)|00\ldots 0 \rangle, \nonumber
\label{}
\end{eqnarray}
indicating that the dynamical properties of the device $H_{1}$ can be approximately expressed as the one of the LMG model 
\begin{equation}
H_{\text{LMG}}=- (J/N)(\mathcal{S}^{z})^2+\mu\mathcal{S}^x.\nonumber \label{}
\end{equation}
The location of the DPT critical point of the LMG model is $\mu_{c}=|J|/2$, leading to $h^{x}_{c}=N\lambda/4$. Note that we only roughly estimate the location of the dynamical critical point of the LMG model. The numerical simulations in the main text are based on the Hamiltonian of the quantum simulator described by Eq.~(1) in the main text.

\begin{acknowledgments}
We thank Chao Song, Qiujiang Guo, Zhen Wang, and Xu Zhang for technical support. Devices were made at the Nanofabrication Facilities at the Institute of Physics in Beijing and National Centre for Nanoscience and Technology in Beijing. The experiment was performed on the quantum computing platform at Zhejiang University. This work was supported by the National Basic Research Program of China (Grants No.~2016YFA0302104, No.~2016YFA0300600 and No.~2017YFA0304300), the National Natural Science Foundations of China (Grants No.~11934018, No.~11725419, No.~11434008, and No.~11904393),
the Strategic Priority Research Program of Chinese Academy of Sciences (Grant No.~XDB28000000), the China Postdoctoral Science Foundation (Grant No.~2018M640055), the AFOSR (Grant No.~FA9550-14-1-0040), the ARO (Grant No.~W911NF-18-1-0358), the JST  Q-LEAP program, the JST CREST (Grant No.~JPMJCR1676), the JSPS-RFBR (Grant No.~17-52-50023, the JSPS-FWO (Grant No.~VS.059.18N), the JSPS Postdoctoral Fellowship (Grant No.~P19326),  the RIKEN-AIST Challenge Research Fund, FQXi,
and the NTT PHI Lab.
\end{acknowledgments}

\cleardoublepage

\emph{Supplementary material for `Probing dynamical phase transitions with a superconducting quantum simulator'}

\noindent \textbf{Device parameters.} ~In Table~\ref{table1}, we present the characteristics for the quantum simulator involving 16 out of the 20 qubits, with XY-control lines, which have been relabelled in the experiments.

\begin{table*}[!htb]
	\centering
	\begin{tabular}{cccccccccccccc}
		%\centering
		\hline
		\hline
		&~~$\omega_{j}/2\pi$ ~&~$T_{1,j}$ ~&~$T_{2,j}^*$ ~&~$g_j/2\pi$ ~&~$\omega_j^r/2\pi$ ~&~$\omega_{j}^{m}/2\pi$~&~~~~$F_{0,j}$~~~~&~~~~$F_{1,j}$~\\
		&(GHz)&($\mu$s)&($\mu$s)&(MHz)&(GHz)&(GHz)&&\\
		\hline
		$Q_1$      & 4.835 & 33 &1.7 & 27.6 & 6.768 & 4.815 & 0.979 & 0.928\\
		$Q_2$      & 5.290 & 21 &1.8 & 27.4 & 6.741 & 5.292 & 0.970 & 0.913\\
		$Q_3$      & 5.330 & 37 &1.8 & 29.1 & 6.707 & 5.350 & 0.978 & 0.920\\
		$Q_4$      & 4.460 & 36 &2.0 & 26.5 & 6.649 & 4.420 & 0.953 & 0.907\\
		$Q_5$      & 4.791 & 32 &2.8 & 29.2 & 6.611 & 4.792 & 0.980 & 0.893\\
		$Q_6$      & 4.870 & 30 &2.1 & 30.1 & 6.558 & 4.850 & 0.989 & 0.938\\
		$Q_7$      & 4.569 & 25 &2.2 & 24.1 & 6.551 & 4.450 & 0.980 & 0.933\\
		$Q_{8}$  & 5.250 & 31 &2.0 & 27.7 & 6.513 & 5.245 & 0.978 & 0.925\\
		$Q_{9}$  & 4.680 & 23 &2.4 & 27.3 & 6.524 & 4.765 & 0.967 & 0.926\\
		$Q_{10}$  & 4.960 & 23 &1.5 & 26.9 & 6.550 & 4.735 & 0.972 & 0.946\\
		$Q_{11}$  & 4.899 & 32 &1.6 & 29.1 & 6.568 & 4.880 & 0.985 & 0.924\\
		$Q_{12}$  & 5.176 & 22 &2.0 & 26.3 & 6.640 & 4.310 & 0.993 & 0.941\\
		$Q_{13}$  & 5.220 & 34 &2.0 & 26.5 & 6.659 & 5.205 & 0.987 & 0.942\\
		$Q_{14}$  & 4.490 & 43 &0.9 & 29.0 & 6.712 & 4.583 & 0.976 & 0.923\\
		$Q_{15}$  & 4.415 & $>$30 &1.9 & 24.6 & 6.788 & 4.375 & 0.967 & 0.944\\
		$Q_{16}$  & 4.766 & 37 &1.5 & 27.5 & 6.758 & 4.906 & 0.970 & 0.945\\
		\hline
		\hline
	\end{tabular}
	\caption{\textbf{Quantum simulator characteristics.} Here, $\omega_{j}$ is the idle frequency of $Q_j$, where single-qubit rotation pulses are applied. $T_{1,j}$ and $T_{2,j}^*$ are the energy relaxation time and Ramsey dephasing time (Gaussian decay) of $Q_j$, respectively, which are measured at the interacting frequency $\omega_I$ ($=\omega_{\mathcal{R}}+\Delta$); $g_j$ denotes the coupling strength between $Q_j$ and the resonator bus $\mathcal{R}$; $\omega_j^r$ is the resonant frequency of $Q_j$'s readout resonator; $\omega_{j}^{m}$ is the resonant frequency of $Q_j$ at the beginning of the measurement process, when its readout resonator is pumped with microwave pulse. Finally, $F_{0,j}$ ($F_{1,j}$) is the probability of detecting $Q_j$ in the $\vert 0\rangle$ ($\vert 1\rangle$) state, when it is prepared in the $\vert 0\rangle$ ($\vert 1\rangle$) state.}\label{table1}
\end{table*}

~

\begin{figure}[!htb]
	\centering
	\includegraphics[width=0.99\linewidth]{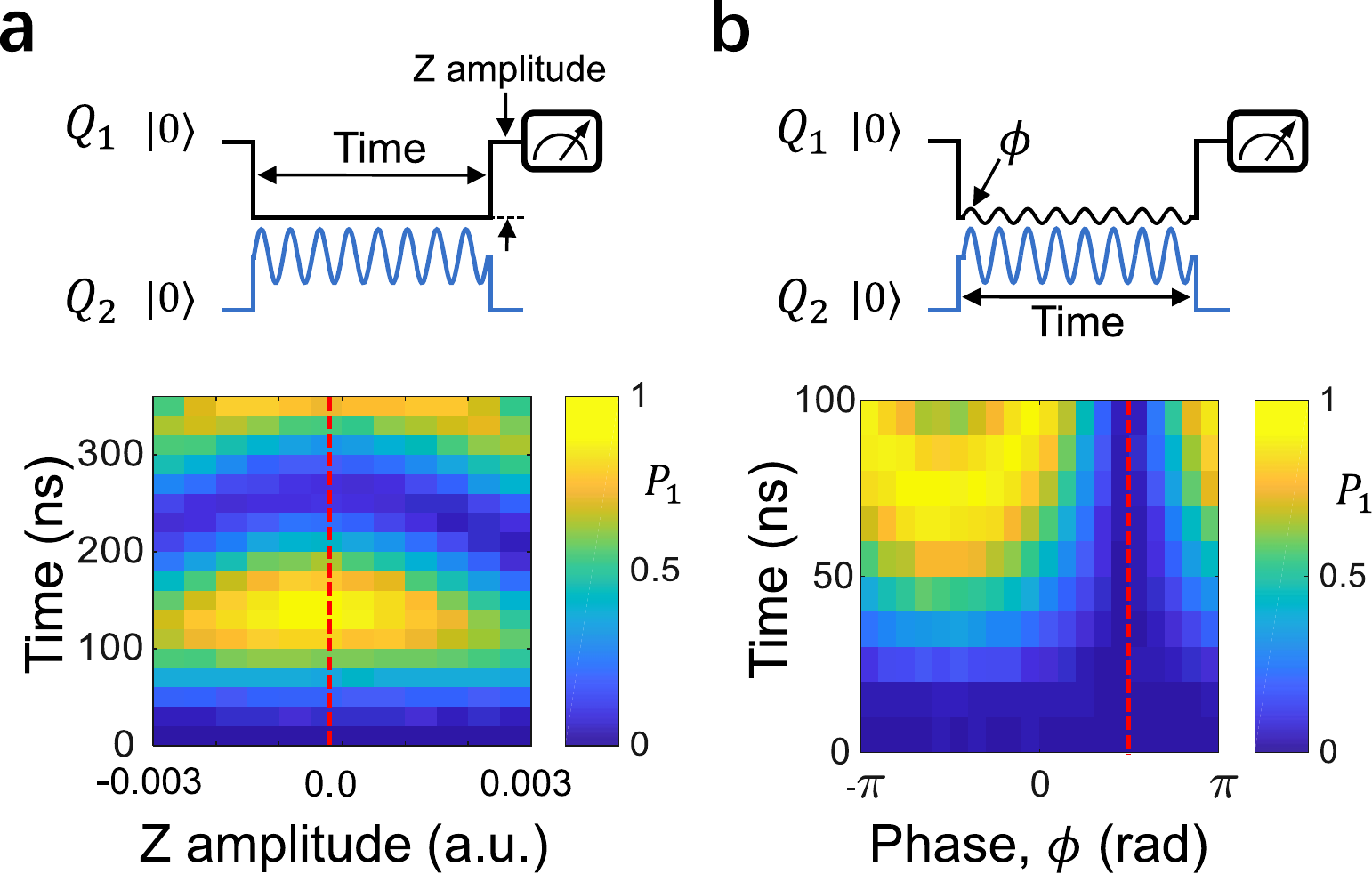}
	\caption{\textbf{Quantification of XY-crosstalk effects.} \textbf{a}, Experimental sequence and results for measuring the XY-crosstalk amplitude. After tuning $Q_1$ to the interacting point $\omega_I$, we apply a strong microwave drive ($h^x_2/2\pi\simeq $15~MHz) on $Q_2$'s microwave drive line with resonant frequency  $\omega_I$. The strong drive will generate a crosstalk Rabi oscillation on $Q_1$. We measure the Rabi oscillations for different  values of $\delta_1$, among which the one with the slowest Rabi oscillation characterises the crosstalk amplitude, as shown by the red dotted line. \textbf{b}, Experimental sequence and results for the measurement of the XY-crosstalk phase. In our experiments, we add a microwave drive on $Q_1$'s XY-control line with an adjustable phase $\phi$. The selection of $\phi$ can induce an enhancement or  neutralisation effect (red dotted vertical line) on $Q_1$'s Rabi oscillation, which can help us identify the XY-crosstalk phase.}\label{e_fig2}
\end{figure}

\noindent \textbf{Correction of XY crosstalk.} The characterisation of the Z-crosstalk effect and its correction have been described in Ref.~3, which are  also the same as in this experiment. In addition, the XY-drive crosstalks between qubits must be corrected, as all qubits are driven by identical microwave drives to quench the system. Different from the Z-crosstalk effects, the characterisation of the XY-crosstalk effects includes the phase calibration of the microwave drives. Fig. \ref{e_fig2} shows the calibration process taking the measurement of the XY crosstalk effect of $Q_2$ to $Q_1$ as an example.

In Fig. \ref{e_fig2}a, to characterise the crosstalk amplitude, $Q_1$ is biased to the interacting frequency $\omega_I/2\pi$, while $Q_2$'s frequency is tuned to a nearby one, e.g., $\omega_I/2\pi-0.08$~GHz (other qubits are decoupled from $Q_1$ and $Q_2$ by tuning their resonant frequencies far away). We apply a strong flattop-envelop microwave pulse with  frequency  $\omega_I$ to $Q_2$'s XY-control line, generating a crosstalk excitation on $Q_1$.
We then monitor the evolution of $Q_1$'s excitations for different resonant frequencies $(\omega_I-\delta_{1})$ of $Q_1$ ($\delta_{1}$ is a small deviation). By fitting the Rabi oscillation of the measured excited probabilities of $Q_1$, we can obtain the crosstalk amplitude of $Q_2$ to $Q_1$.

In Fig. \ref{e_fig2}b, we characterise the crosstalk phase of $Q_2$ to $Q_1$, with the same frequency arrangement as that in Fig. \ref{e_fig2}a. However, to cancel the crosstalk effect of $Q_2$ to $Q_1$, a microwave pulse on $Q_1$'s XY-control line, with  amplitude equal to the crosstalk amplitude of $Q_2$ to $Q_1$, should also be added. In this process, we monitor the evolution of $Q_1$'s excitations for different phases of the microwave pulses on $Q_2$'s XY-control line, while fixing the microwave phase of $Q_1$ to zero. The excitations of $Q_1$ are almost completely inhibited during the whole evolution process at a specific phase, as can be seen from Fig. \ref{e_fig2}b, showing that the phase difference of the microwave drives on the XY-control lines of $Q_1$ and $Q_2$ is $\pi$.

Calibrations of other pairs of qubits are performed with a similar method. After quantifying these crosstalk effects, we correct these issues in experiments by considering the case which only involves two qubits $Q_1$ and $Q_2$. We bias these two qubits to $\omega_I/2\pi$ and simultaneously apply resonant microwave pulses on their XY-control lines with  amplitudes  $Ae^{i\phi_A}$ and $Be^{i\phi_B}$ for $Q_1$ and $Q_2$, respectively (other qubits are decoupled from $Q_1$ and $Q_2$ by tuning their resonant frequencies far away). Here $\phi_A$ and $\phi_B$ represent the microwave phases of $Q_1$ and $Q_2$, respectively. If no correction is made, the microwave amplitude and phase of each qubit can be represented as
\begin{equation}
{
	\left[ \begin{array}{c}
	Ae^{i\phi_A}+a_{12}Be^{i(\phi_{12}+\phi_B)}\\
	a_{21}Be^{i(\phi_{21}+\phi_B)}+Be^{i\phi_B}
	\end{array}
	\right ]}=\tilde{M}_{xy}^{Q_1,Q_2}
{\left[ \begin{array}{c}
	Ae^{i\phi_A}\\
	Be^{i\phi_B}
	\end{array}
	\right ]},\nonumber
\end{equation}
where
\begin{equation} 	\tilde{M}_{xy}^{Q_1,Q_2}
=\left[%^{~~~~~~~~1}_{ a_{21}\exp({i\phi_{21}})}{~}^{ a_{12}\exp({i\phi_{12}})}_{~~~~~~~~1}\right]
\begin{array}{cc}1 & a_{12}e^{i\phi_{12}} \\a_{21}e^{i\phi_{21}} & 1\end{array}\right]\nonumber
\end{equation} 
is the XY-crosstalk matrix measured with the technique described above. On the contrary, if we set the microwave amplitudes and phases of the qubits to $A'e^{i\phi_A'}$ and $B'e^{i\phi_B'}$  in advance, the microwaves we apply to the XY-control lines should be corrected as
\begin{equation}
(\tilde{M}_{xy}^{Q_1,Q_2})^{-1}
{\left[ \begin{array}{c}
	A'e^{i\phi_A'}\\
	B'e^{i\phi_B'}
	\end{array}
	\right ]}\;.\nonumber
\end{equation}

\begin{figure}[!htb]
	\centering
	\includegraphics[width=0.99\linewidth]{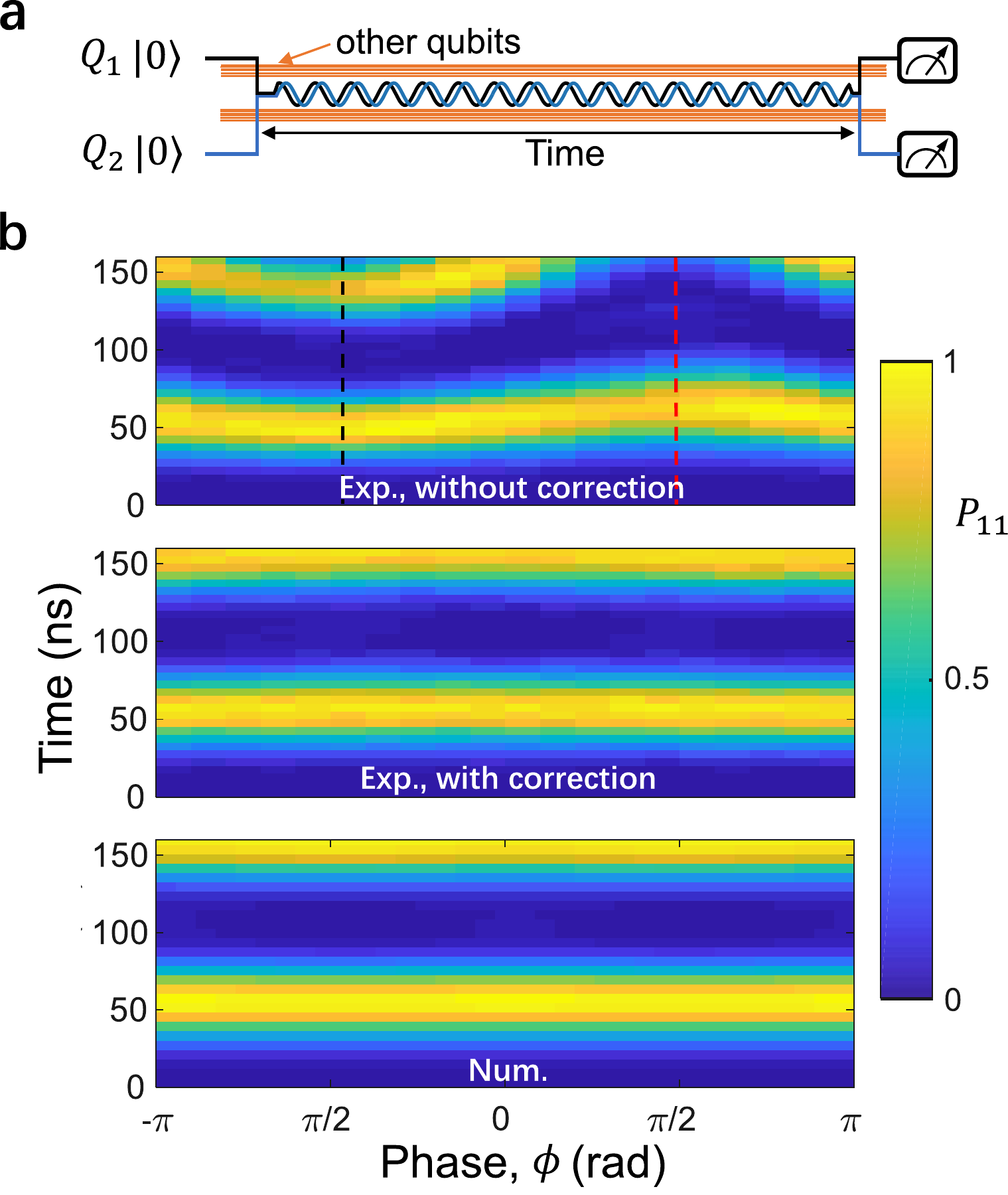}
	\caption{\textbf{Experimental test of the XY-crosstalk correction.} \textbf{a}, The experimental sequence. We tune  two qubits on resonance at $\omega_I$, while other qubits are arranged in the vicinity, and apply resonant microwave drives ($h^x/2\pi\simeq$5~MHz for each qubit) on these two qubits's XY-control lines with a controllable phase difference of $\phi$. \textbf{b}, The measured probabilities $P_{11}$ of the $\vert 11\rangle$-state,  versus $t$ and $\phi$, in cases with and without applying the XY-crosstalk correction, compared with the numerical results. When no XY-crosstalk correction is made, the measured oscillation periods of $P_{11}$ for different $\phi$ values show an obvious inconsistency, indicating an enhancement (black dotted vertical line) or neutralisation (red dotted vertical line) effects induced by the XY crosstalk. After applying the XY-crosstalk correction, the experimental results are in good agreement with the numerical results.}\label{e_fig3}
\end{figure}

We have performed experiments to verify the validity of this XY-crosstalk correction, where we bias $Q_1$ and $Q_2$ to $\omega_I$ to open the interaction and simultaneously apply identical microwaves to these two qubits. The results are displayed in Fig. \ref{e_fig3}, demonstrating the validity of this correction. The same treatment can be easily extended to the multi-qubit case.

Note that in the experiments we apply two kinds of microwave drives on the qubits. One establishes the transverse field with the same driving frequency $\omega_I$ for all qubits, and the other is the rotation pulse applied at each qubit's idle frequency. The XY-crosstalk correction is only applied when we impose microwave drives on all the qubits, to generate the identical transverse fields. As for the rotation pulses, the XY-crosstalk effects are negligible due to the large detuning between the idle points of qubits.

~

\begin{figure}[!htb]\centering
	\includegraphics[width=0.99\linewidth]{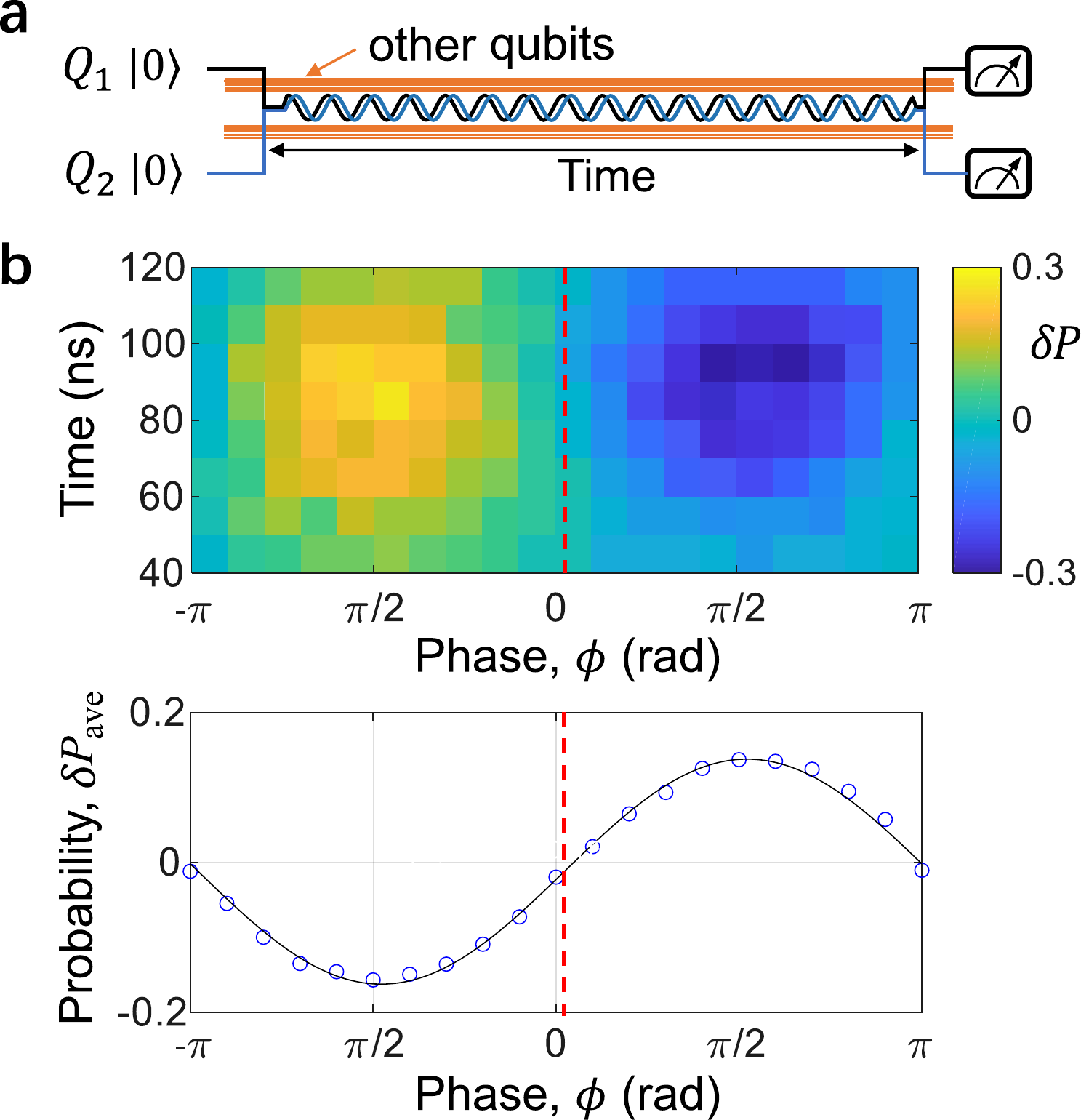}
	\caption{\textbf{Phase alignment of the transverse field.} \textbf{a}, Experimental sequence. Two qubits ($Q_1$ and $Q_2$) are detuned from the resonator bus $\mathcal{R}$ by the same amount $\Delta/2\pi\simeq-450$~MHz, while other qubits are arranged in the vicinity of this point to minimise Z-crosstalk effects. We then apply resonant microwave drives on these two qubits with the same magnitude ($h^x/2\pi\simeq2$~MHz) but a phase difference of $\phi$ and monitor the dynamics from 40~ns to 120~ns by recording the probabilities of $Q_1$ and $Q_2$, denoted as $P_1^{Q_1}$ and $P_1^{Q_2}$. \textbf{b}, 2D graph of $\delta\! P\equiv P_1^{Q_2}-P_1^{Q_1}$ as a function of $t$ and $\phi$ (top) and the time-averaged $\delta\! P$ (bottom). We fit this curve with a sine function to extract the phase offset (red dotted vertical line), which will be added to the microwave drive of $Q_2$ to ensure the phase alignment between these two qubits.}\label{e_fig4}
\end{figure}

\noindent \textbf{Calibration of the transverse field.} The term of the local transverse field $h^{x}_{j}\sigma^x_j$ for each $Q_j$ in Eq.~(1) in the main text is enabled by the resonant microwave drives with calibrated magnitude and phase. To ensure the uniformity of the driving magnitude, we perform  Rabi oscillation measurements on each stand-alone qubit $Q_j$ at the interacting frequency $\omega_I$. The qubit exposed to the resonant microwave drive will experience an oscillation of its excited-state probability, where the oscillation frequency $h^{x}_{j}/\pi$ can be adjusted by the driving  amplitude. For the phase calibration {of the transverse field}, when applying microwave drives with a flattop envelop to each $Q_j$, we actually obtain the form $h^{x}_j(e^{-i\phi_j}\sigma_{j}^{+}+e^{i\phi_j}\sigma_{j}^{-})$, where $\phi_j$ is the microwave phase sensed by each $Q_j$ and may be different from each other, because of the length disparities between each $Q_j$'s XY-control wires. The experiments require the uniformity of $\phi_j$, which can be achieved by the following calibration process. Here, we consider two qubits ($Q_1$ and $Q_j$), equally detuned from the resonator bus by $\Delta/2\pi\simeq-450$ MHz and driven by resonant microwaves through their own XY-control lines with the driving phases of these two qubits set to 0 and $\phi_j,$ respectively. The two-qubit Hamiltonian can be written as
\begin{equation}
\begin{split}
H_{1j} /\hbar = &\lambda_{1j}(\sigma^-_1\sigma^+_j+\sigma^-_j\sigma^+_1) \\
&+h^{x}(\sigma^x_1+e^{-i\phi_j}\sigma^+_j+e^{i\phi_j}\sigma^-_j), \label{eqs2}
\end{split}\nonumber
\end{equation}
where $\lambda_{1j}$ is the coupling strength between $Q_1$ and $Q_j$, and $h^x$ represents the driving magnitude on the two qubits. In experiments, we start with the ground state and monitor the evolution of the two-qubit system under the above Hamiltonian for different  values of $\phi_j$. We select $Q_1$ as the reference and adjust the $\phi_j$ of other qubits to make them pairwise aligned with that of $Q_1$. Note that when performing the phase check of $Q_1$ and $Q_j$ at the interacting point $\omega_I$,  the frequencies of other qubits are arranged in the vicinity (about 50 to 100~MHz away from $\omega_I$) to minimise the Z-crosstalk effect. The calibration sequence and experimental results for different $\phi_j$ are displayed in Fig. \ref{e_fig4}.

~

\noindent \textbf{Phase calibration of the rotation pulse.} As can be seen from Fig.~1c, after the evolution under the quenched Hamiltonian, we apply the rotation pulse on each qubit before the joint readout to measure the physical quantities,  including the average spin magnetisations $\langle\sigma^{x}(t)\rangle$ and $\langle\sigma^{y}(t)\rangle$. The rotation operation on each qubit is realised by a Gaussian-envelope microwave pulse with a full width at half maxima of 20~ns, which has been characterised by randomised benchmarking with a fidelity above 0.99 for both $X_{\pi/2}$ and $Y_{\pi/2}$ rotation gates.

\begin{figure}[!htb]\centering
	\includegraphics[width=0.99\linewidth]{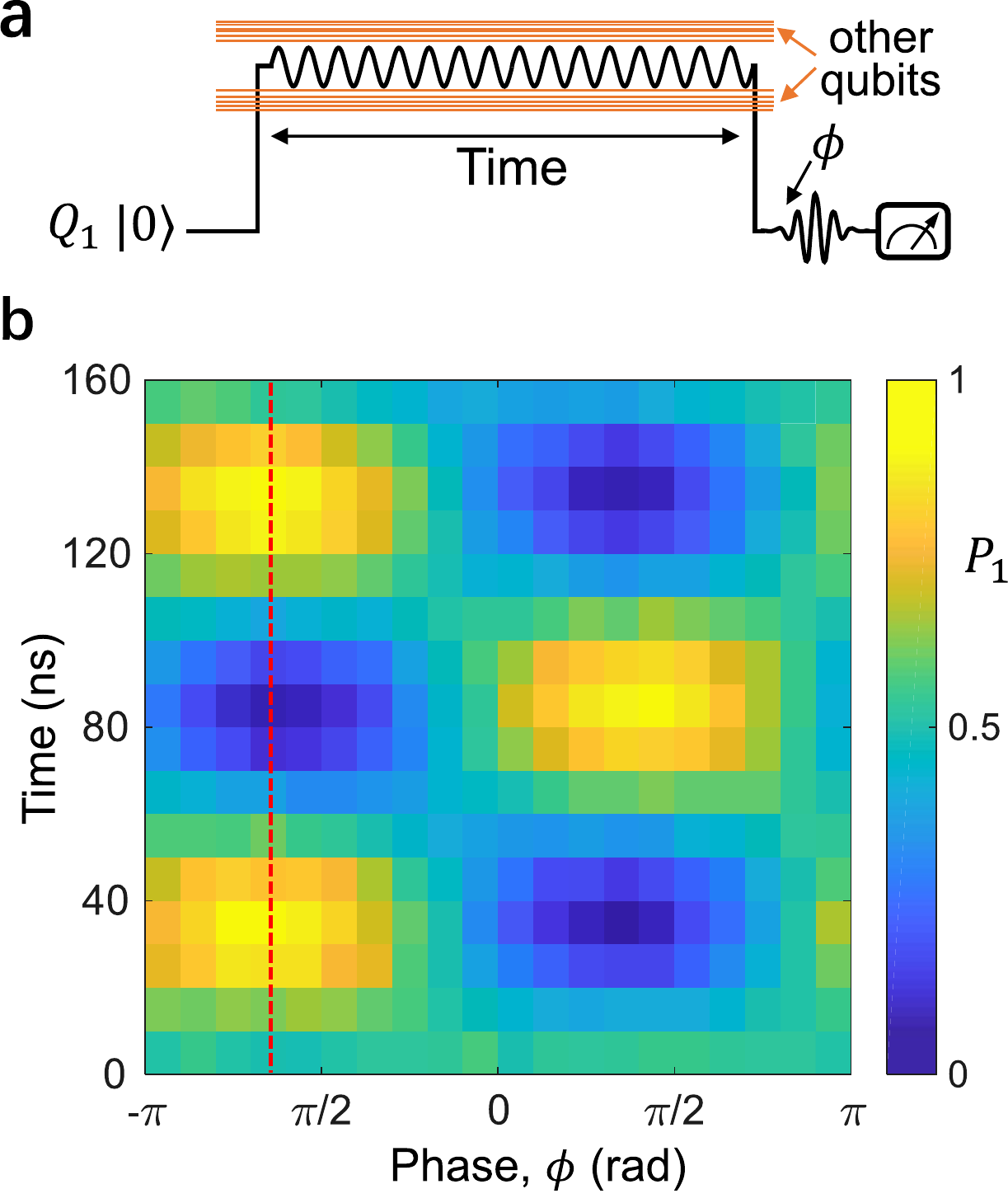}
	\caption{\textbf{Phase alignment of the rotation pulse.} \textbf{a}, Experimental sequence taking $Q_1$ as an example. The qubit $Q_1$ is detuned from the resonator bus $\mathcal{R}$ by about $\Delta/2\pi\simeq-450$~MHz, while other qubits are arranged in the vicinity of this point to minimise Z-crosstalk effects. Simultaneously, we apply on $Q_1$  resonant microwave drives with a magnitude of $h^x/2\pi\simeq5$~MHz, after which we quickly bias $Q_1$ to its idle point and apply a rotation pulse with a specific phase $\phi$ before the readout. \textbf{b}, 2D graph of the measured excited probabilities $P_1$ as a function of $t$ and $\phi$. The phase offset pointed by the  red dotted vertical line should be added to $Q_1$'s rotation pulse to align the phases.}\label{e_fig5}
\end{figure}

To mainly compensate for the dynamic phase caused by frequency tuning through the sequence, the phase of each rotation pulse needs to be corrected. The calibration process is presented in Fig. \ref{e_fig5}a, taking $Q_1$ as an example. The calibrated qubit is biased to the interacting frequency $\omega_I$ with a rectangular pulse, while the frequencies of other qubits are arranged in the vicinity to minimise the Z-crosstalk effect. Almost simultaneously, $Q_1$ is driven by a flattop-envelope microwave pulse with  frequency  $\omega_I$. Then, we bias  $Q_1$ back to its idle frequency and apply a $\pi/2$-rotation pulse before the readout. We record the probabilities of $Q_1$'s excited state during this dynamics for different phases $\phi$ of the rotation pulse. The results are displayed in Fig. \ref{e_fig5}b, where the phase offset used for the correction is highlighted by the  red dotted vertical line.

~

\noindent \textbf{Measurement of the spin-squeezing parameter.} The calculation of the spin-squeezing parameter $\xi^{2}$ consists of the following steps. The first step is to calculate the mean-spin direction $\vec{n}_{0}=(\sin\theta\cos\phi,\sin\theta\sin\phi,\cos\theta)$, where
\begin{eqnarray}
\theta=\arccos\left(\frac{\mathcal{S}^{z}}{|\vec{\mathcal{S}}|}\right), \nonumber
\label{theta}
\end{eqnarray}
and
\begin{eqnarray}
\phi=\left\{
\begin{array}{lcl}
\arccos\left(\frac{\langle\mathcal{S}^{x}\rangle}{|\vec{\mathcal{S}}|}\right) & & \text{if}\indent{\langle\mathcal{S}^{y}\rangle>0}\\
2\pi - \arccos\left(\frac{\langle\mathcal{S}^{x}\rangle}{|\vec{\mathcal{S}}|}\right) & & \text{if}\indent{\langle\mathcal{S}^{y}\rangle<0} \nonumber
\end{array} \right.,
\label{phi}
\end{eqnarray}
with $|\vec{\mathcal{S}}|^{2} = \langle\mathcal{S}^{x}\rangle^{2} + \langle\mathcal{S}^{y}\rangle^{2} + \langle\mathcal{S}^{z}\rangle^{2}$.
The second step is to obtain the expression of $\mathcal{S}_{\vec{n}_{\perp}}$ and to {minimize its variance.} We can obtain two orthogonal bases, $\vec{n}_{1}=(-\sin\phi,\cos\phi,0)$ and $\vec{n}_{2} = (\cos\theta\cos\phi,\cos\theta\sin\phi,-\sin\theta)$, perpendicular to $\vec{n}_{0}$. Then, $\mathcal{S}_{\vec{n}_{\perp}}$ can be represented as $\vec{\mathcal{S}}\cdot\vec{n}_{\perp}$, with 
%\begin{eqnarray}
$\vec{n}_{\perp}=\vec{n}_{1}\cos\vartheta + \vec{n}_{2}\sin\vartheta$ %\nonumber
%\label{n_perp}
%\end{eqnarray}
and $\vartheta\in [0,2\pi]$. The minimum in Eq.~(2) of the main text is actually equivalent to the optimisation of $\vartheta$. {It turns out that} the optimum procedure finally gives an elegant formula
\begin{eqnarray}
\begin{split}
\xi^{2} = &\frac{2}{N}[\langle(\mathcal{S}^{\vec{n}_{1}})^2+(\mathcal{S}^{\vec{n}_{2}})^2\rangle \\
&+\sqrt{\langle(\mathcal{S}^{\vec{n}_{1}})^2-(\mathcal{S}^{\vec{n}_{2}})^2\rangle^{2}+\langle\{\mathcal{S}^{\vec{n}_{1}},\mathcal{S}^{\vec{n}_{2}}\}\rangle^{2}}],
\label{cal_ss}
\end{split}
\end{eqnarray}
with $\{\mathcal{S}^{\vec{n}_{1}},\mathcal{S}^{\vec{n}_{2}}\}=\mathcal{S}^{\vec{n}_{1}}\mathcal{S}^{\vec{n}_{2}} + \mathcal{S}^{\vec{n}_{2}}\mathcal{S}^{\vec{n}_{1}}$.

\begin{figure}[]\centering
	\includegraphics[width=0.99\linewidth]{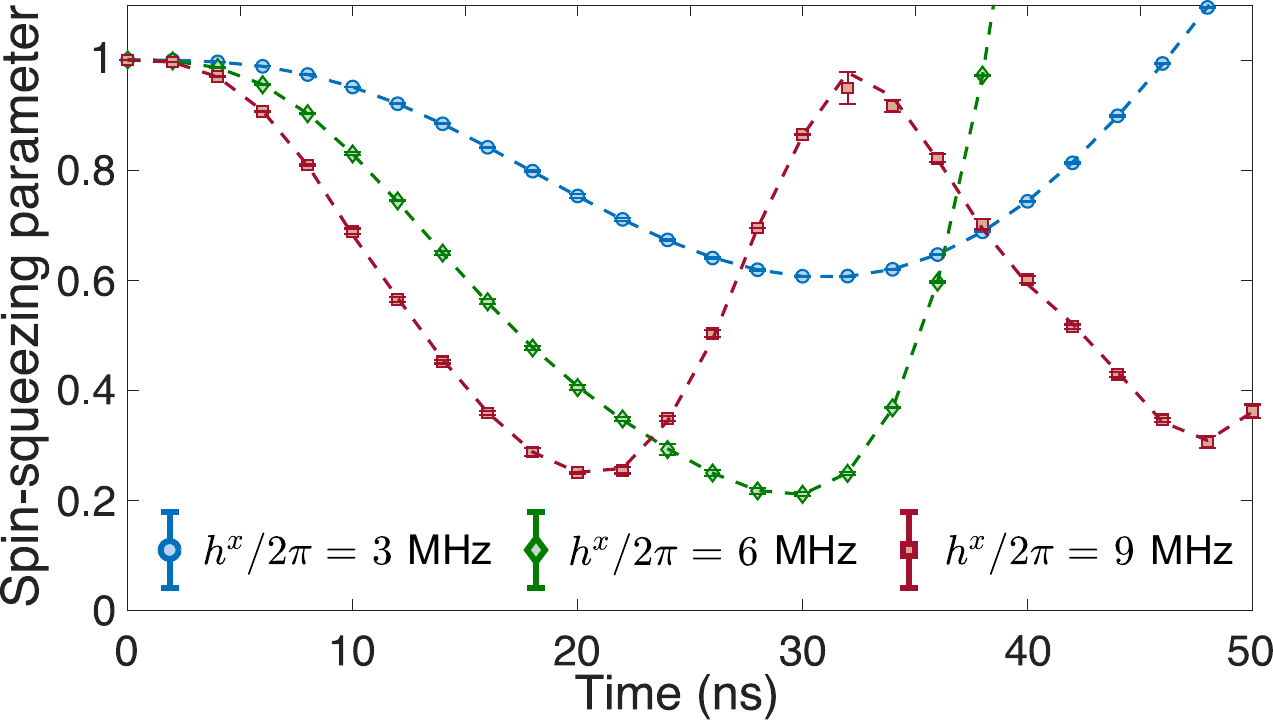}
	\caption{\textbf{Numerical calculation of the spin-squeezing parameter.} The dashed curves are the strict results according to Eq.~(\ref{cal_ss}), while the points are calculated with the method described above. To estimate the error bar, we repeat the calculation 5 times for different values of $t$ and $h^x$. For each time, we randomly select 5 groups of $\{G_1^i,G_2^i\}$ and average the results.}\label{e_fig6}
\end{figure}

{We measure $\langle\mathcal({S}^{\vec{n}_{1}})^2\rangle$ and $\langle\mathcal({S}^{\vec{n}_{2}})^2\rangle$}
by applying single-qubit rotations to move the $\vec{n}_{1}$ ($\vec{n}_{2}$) axis in the Bloch sphere to the $z$-axis before readout. For $\langle\{\mathcal{S}^{\vec{n}_{1}},\mathcal{S}^{\vec{n}_{2}}\}\rangle$, it boils down to the measurement of two-spin correlators
\begin{eqnarray}
\langle\{\mathcal{S}^{\vec{n}_{1}},\mathcal{S}^{\vec{n}_{2}}\}\rangle = \frac{1}{4}(\sum_{i\neq j} \langle\sigma_{i}^{\vec{n_{1}}}\sigma_{j}^{\vec{n_{2}}}\rangle + \sum_{i\neq j}\langle\sigma_{i}^{\vec{n_{2}}}\sigma_{j}^{\vec{n_{1}}}\rangle). \nonumber
\label{}
\end{eqnarray}
{To characterise the two spin correlators for all combinations (16 $\times$ 15 $\times$ 2),} we adopt the following methods: First, we divide the 16 qubits into 2 groups randomly, e.g., $G_1^1=$\{$Q_1$--$Q_8$\} and $G_2^1=$\{$Q_9$--$Q_{16}$\}. Next, we apply rotation pulses on the qubits in $G_1^1$ to bring the $\vec{n}_1$-axis to the $z$-axis, and simultaneously apply other rotation pulses on qubits in $G_2^1$ to bring the $\vec{n}_2$-axis to the $z$-axis, after which the 16-qubit joint readout is executed, yielding the probabilities $P$=\{$P_{00...0}$, $P_{00...1}$, ..., $P_{11...1}$\}. Finally, by calculating $\sum_{j=1}^{2^{16}}P_{j}\mathcal{S}^{zz(G_1^1,G_2^1)}_{j,j}$, with $\mathcal{S}^{zz(G_1^1,G_2^1)}$ written as
\begin{equation}
\mathcal{S}^{zz(G_1^1,G_2^1)}=\sum_{i\in G_1^1}^{}\sigma^{z}_i\sum_{j \in G_2^1}^{}\sigma^{z}_j, \nonumber
\end{equation}
we obtain the summation of two-spin correlators for $8\times8=64$ combinations ($Q_1$--$Q_9$, $Q_1$--$Q_{10}$, ..., $Q_1$--$Q_{16}$, $Q_2$--$Q_{9}$, ..., $Q_{8}$--$Q_{16}$), i.e., 
\begin{equation}
P_{n_1n_2}^1(G_1^1,G_2^1)=\sum_{i\in G_1^1, j\in G_2^1}^{}\langle\sigma^{\vec{n}_1}_i\sigma^{\vec{n}_2}_j\rangle. \nonumber
\end{equation}
Moreover, by exchanging the rotation pulses applied to qubits in these two groups, we can obtain
\begin{equation} 
P_{n_2n_1}^1(G_1^1,G_2^1)=\sum_{i\in G_1^1, j\in G_2^1}^{}\langle\sigma^{\vec{n}_2}_i\sigma^{\vec{n}_1}_j\rangle.\nonumber
\end{equation} 
After repeating this process 5 times, where 16 qubits are divided into two different  groups of equal size, we can approach $\langle\{\mathcal{S}^{\vec{n}_{1}},\mathcal{S}^{\vec{n}_{2}}\}\rangle$ by averaging the overall results 
\begin{equation}
\frac{16\times 15}{64\times 5}\sum_{i=1}^{5}[P_{n_1n_2}^i(G_1^i,G_2^i)+P_{n_2n_1}^i(G_1^i,G_2^i)].\nonumber
\end{equation}
This method has been verified by numerical simulations that possess a very high precision, as illustrated in Fig. \ref{e_fig6}.

~

\begin{figure}[!htb]\centering
	\includegraphics[width=0.99\linewidth]{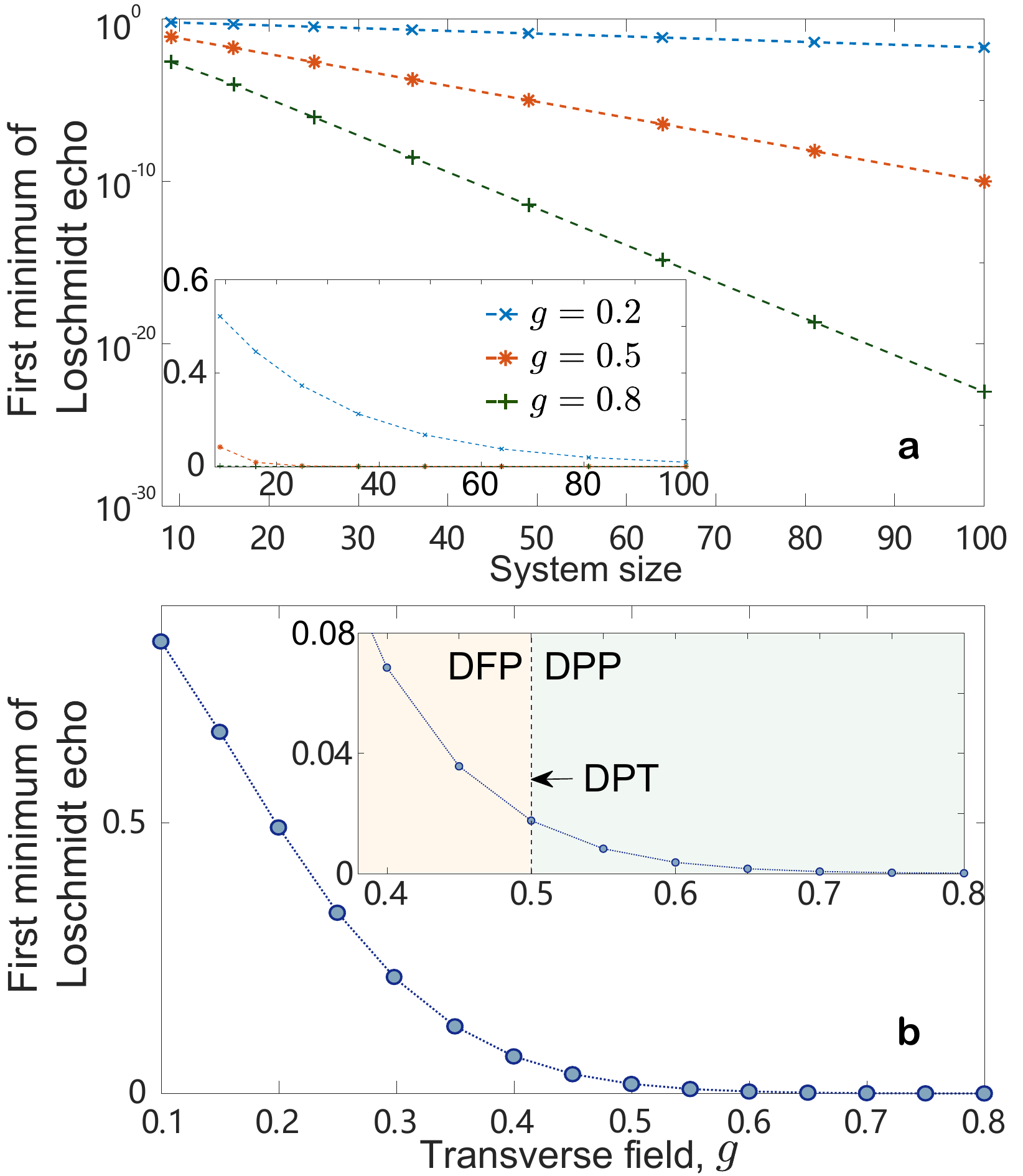}
	\caption{\textbf{Numerical results of the Loschmidt echo in the LMG model.} \textbf{a}, The value of the first minimum of the Loschmidt echo $\mathcal{L}_{\text{min}}^{(1)}$ scales with the system's size $N$. \textbf{b}, The value of $\mathcal{L}_{\text{min}}^{(1)}$ as a function of $g$ for $N=16$.}\label{e_fig8}
\end{figure}

\noindent \textbf{Finite-size effect of the Loschmidt echo in the LMG model.} The results in the main text are in good agreement with the theoretical predictions based on the LMG model.
%$H_{\text{LMG}} = (J/N) (\mathcal{S}^{z})^{2} + g \mathcal{S}^{x}$. 
%Consequently, the numerical results of the LMG model have referenced value for the analysis of experimental data. 
It has been shown that the Loschmidt echo cannot be strictly equal to 0 in a finite-size LMG model~\cite{ref28}. In Fig. \ref{e_fig8}a, we present the first minimum value of the Loschmidt echo $\mathcal{L}_{\text{min}}^{(1)}$ as a function of the LMG model's size $N$ with $J=1$ and different $g$, showing a perimeter law $\mathcal{L}_{\text{min}}^{(1)}\sim\exp(-\alpha N)$, with $\alpha>0$. Although $\mathcal{L}_{\text{min}}^{(1)}\rightarrow 0$ as $N\rightarrow \infty$ for arbitrary $g>0$, we can still observe a drastic difference of $\mathcal{L}_{\text{min}}^{(1)}$ in the two phases (Fig. \ref{e_fig8}b). Based on the above discussions, we believe that for the 16-qubit system, the value of $\mathcal{L}_{\text{min}}^{(1)}$ smaller than $\sim10^{-2}$ can be a characteristic of the dynamical paramagnetic phase.

~

\begin{figure}[!htb]\centering
	\includegraphics[width=0.99\linewidth]{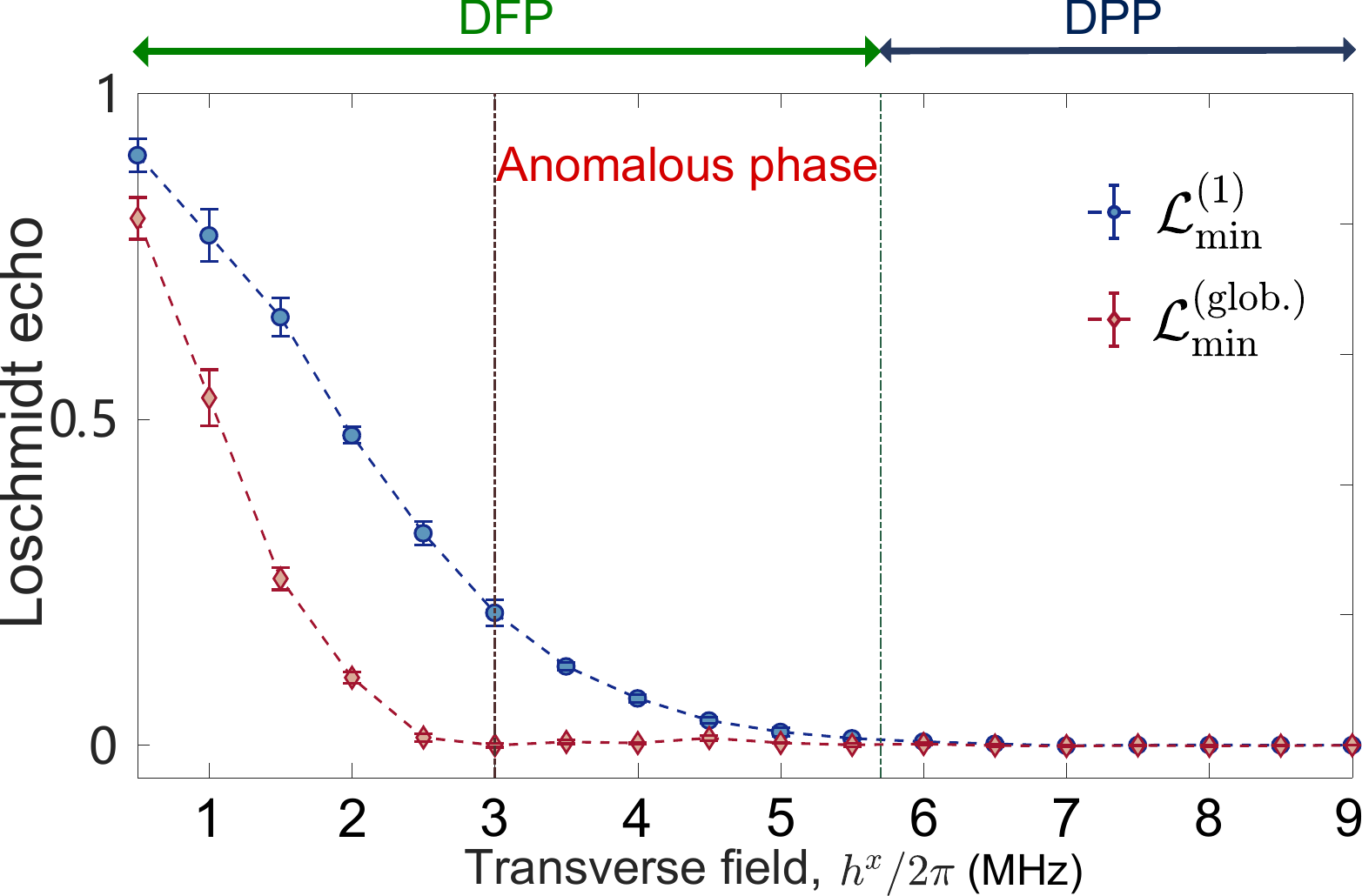}
	\caption{\textbf{Experimental data for the long-time behaviour of the Loschmidt echo.} The value of $\mathcal{L}_{\text{min}}^{(\text{glob.})}$ as a function of $h^{x}$, with $\mathcal{L}_{\text{min}}^{(\text{glob.})}$ referring to the minimum value of the Loschmidt echo during its time evolution with a final time of around 600~ns.}\label{e_fig9}
\end{figure}

\noindent \textbf{Possible signature of the anomalous dynamical phase.} In addition to $\mathcal{L}_{\text{min}}^{(1)}$, including short-time properties of the Loschmidt echo $\mathcal{L}(t)$, the long-time evolution of $\mathcal{L}(t)$  provides more information, such as the signature of the anomalous dynamical phase, i.e., the $\mathcal{L}(t)$ may approach  zero for long times, suggesting the existence of the nonanalytical point of the rate function $r(t)=-N^{-1}\log[\mathcal{L}(t)]$ in the dynamical ferromagnetic phase. The anomalous dynamical phase only exists for  models with long-range interactions, and is absent for  models with short-range interactions. In Fig. \ref{e_fig9}, we present the experimental data of $\mathcal{L}_{\text{min}}^{(1)}$ compared with the global minimum value of $\mathcal{L}(t)$, i.e., $\mathcal{L}_{\text{min.}}^{(\text{glob.})}$, during its dynamics with a final time $t_{f} \simeq 600$~ns, showing a possible signature of the  anomalous dynamical phase, which can enlighten further works for investigating the anomalous dynamical phase.

~

\begin{figure}[!htb]
	\centering
	\includegraphics[width=0.99\linewidth]{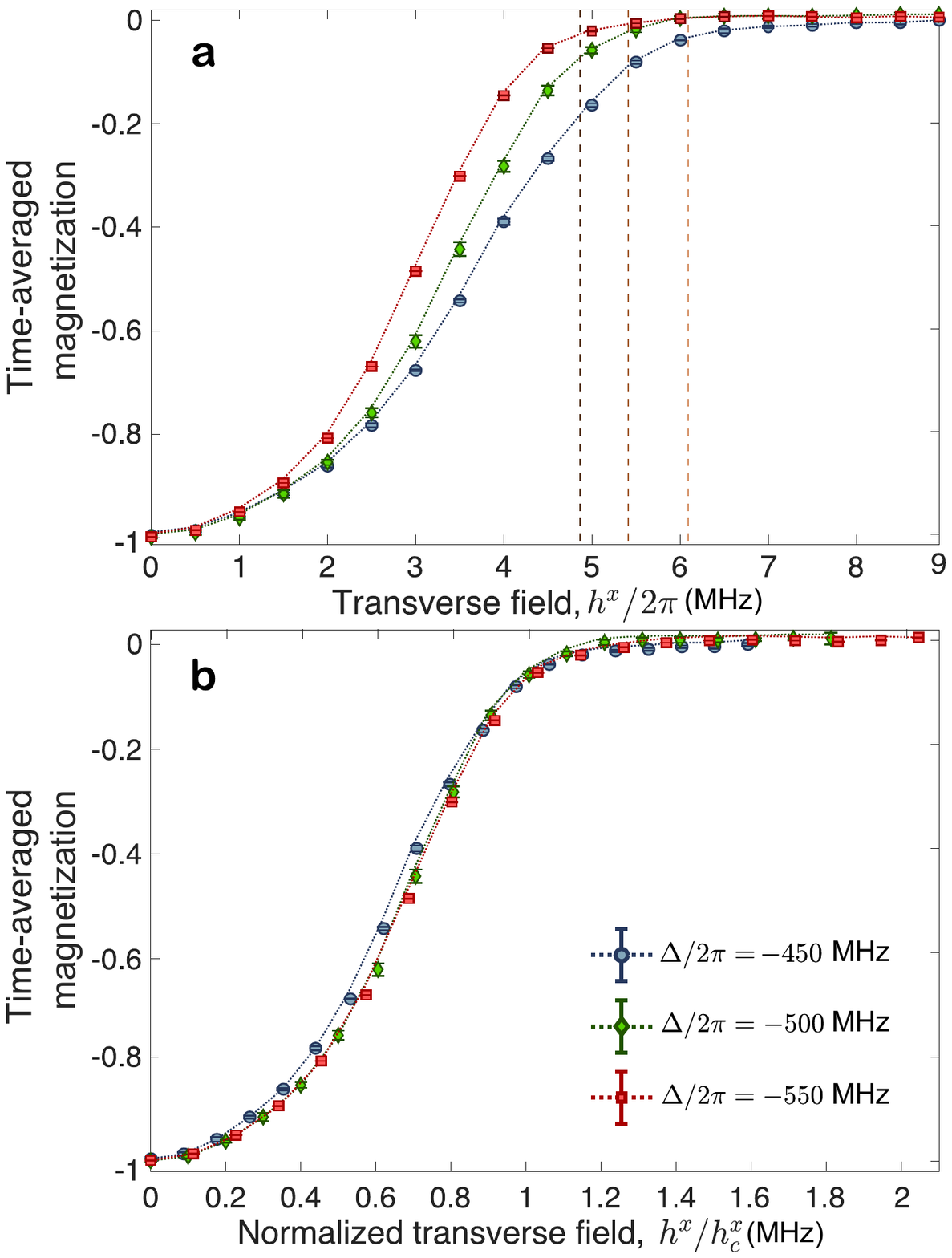}
	\caption{\textbf{Experimental data for non-equilibrium order parameter with  different values of the detuning $\Delta$.} \textbf{a}, The order parameter $\overline{\langle\sigma^{z}\rangle}$ as a function of the field strength $h^{x}$. The theoretically predicted critical points for $\Delta/2\pi\simeq-450$~MHz, $-500$~MHz and $-550$~MHz are $h^{x}_{c}/2\pi \sim 5.7$~MHz, $5.0$~MHz, and $4.4$~MHz, respectively, as highlighted by the dashed vertical lines. \textbf{b}, The same data in \textbf{a} but with $h^{x}$ normalized  by its critical value $h^{x}_{c}$.}\label{e_fig1}
\end{figure}

\begin{figure*}[!htb]\centering
	\includegraphics[width=0.99\linewidth]{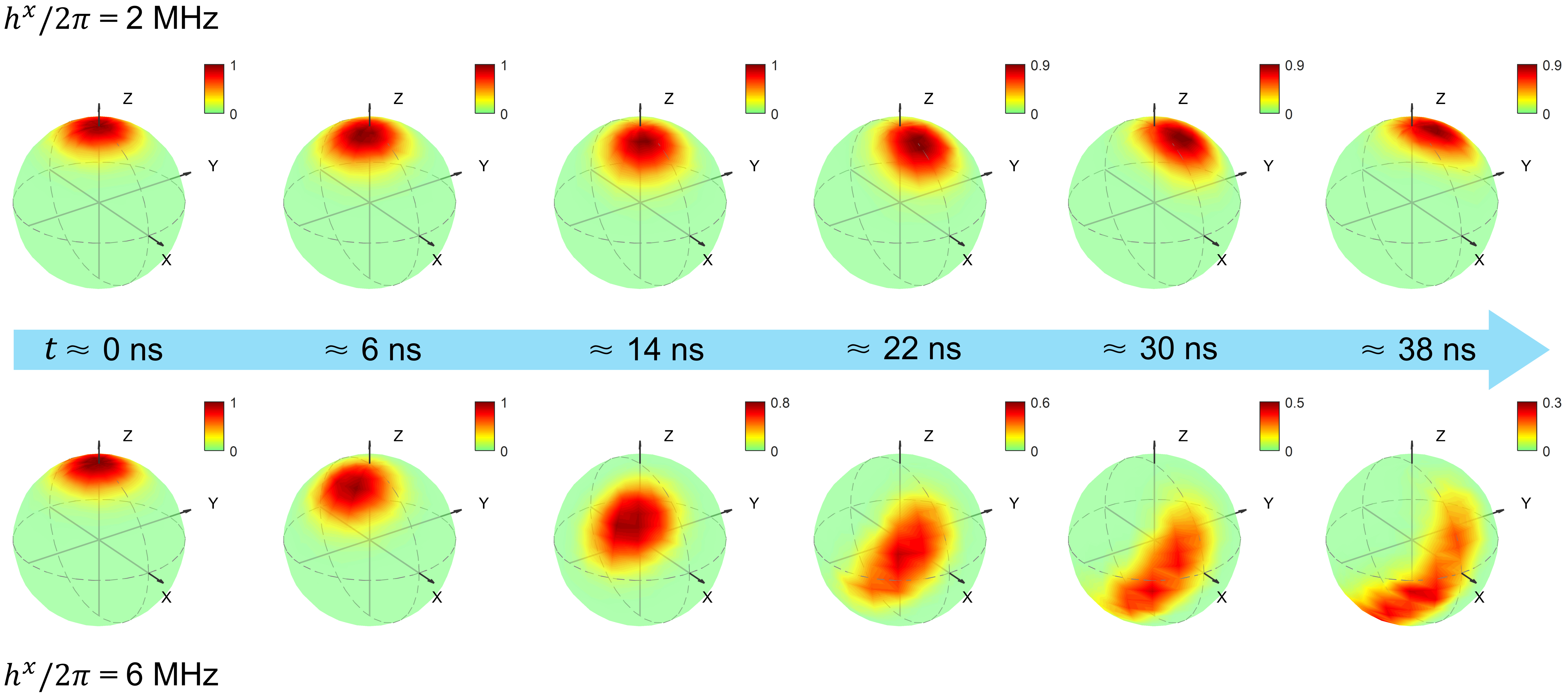}
	\caption{\textbf{Dynamics of the quasidistribution $Q$-function.} The quasidistributions $Q_{\rm exp}(\theta,\phi)$ at different time intervals, for $h^x/2\pi=$2~MHz (up) and 6~MHz (down), respectively.}\label{e_fig7}
\end{figure*}

\noindent \textbf{Additional experimental data.} In Fig.\ref{e_fig1}, we plot the experimentally measured non-equilibrium order parameter as a function of the transverse field magnitude for different values of the detuning $\Delta$.  In Fig.\ref{e_fig7}, we display the evolution of the experimental quasidistribution $Q$-function for two transverse field magnitudes.

\end{document}